\documentclass[aps,prb,twocolumn,groupedaddress, showpacs,showkeys]{revtex4}
\usepackage{graphicx}
\usepackage{amsmath}
\usepackage{amsfonts}
\usepackage{bm}
\usepackage{bbm}
\usepackage{epstopdf}
\usepackage{hyperref}
\usepackage{psfrag}
\usepackage{subfigure}
\usepackage{url}
\usepackage{xspace}
\usepackage{dsfont}
\usepackage{color}
\usepackage{tabularx}
\newcommand{\li}{\langle}
\newcommand{\re}{\rangle}
\newcommand{\f}{\mathbf}
\newcommand{\bra}[1]{\left\langle #1 \right|}    
\newcommand{\ket}[1]{\left| #1 \right\rangle}    
 
\newcommand{\matrixelem}[3]{\left\langle #1 \left| #2 \right| #3 \right\rangle} 
\newcommand{\etal}{\textit{et al.}\xspace}

\newcommand{\eV}{$\mathrm{eV}$\xspace}       
\newcommand{\meV}{$\mathrm{meV}$\xspace}      
\newcommand{\real}{\operatorname{Re}}
\newcommand{\imag}{\operatorname{Im}}
\newcommand{\trace}{\operatorname{Tr}}

\newcommand{\mtrx}[1]{\mathbf{#1}}
\begin{document}

%%%%%%%%%%%%%%%%%%%%%%%%%%%%%%%%%%%%%%%%%%%%%%%%%%%%%%%%%%%
%......Title and other stuff
\title{Theory of band gap bowing of disordered substitutional II-VI and 
       III-V semiconductor alloys}

\author{Daniel Mourad}
\email{dmourad@itp.uni-bremen.de}
\affiliation{Institute for Theoretical Physics,
             University of Bremen,
             28359 Bremen, Germany}
\author{Gerd Czycholl}
\affiliation{Institute for Theoretical Physics,
             University of Bremen,
             28359 Bremen, Germany}

\date{\today}
%============================================================
% Abstract
%============================================================
\begin{abstract}
%\abstract{
For a wide class of technologically relevant compound III-V and
II-VI semiconductor materials AC and BC mixed crystals (alloys) of the type
A$_x$B$_{1-x}$C
can be realized. As the electronic properties like the
bulk band gap vary continuously with $x$, any band gap in between that of the pure
AC and BC systems can be obtained by choosing the appropriate concentration $x$,
granted that the respective ratio is miscible and thermodynamically 
stable.
In most cases the band gap does not vary linearly with $x$, but a pronounced bowing
behavior as a function of the concentration is observed.
In this paper we show that the electronic properties of such A$_x$B$_{1-x}$C
semiconductors and, in particular, the band gap bowing can well be described and
understood starting from
empirical tight binding models for the pure AC and BC systems.
The electronic properties of the A$_x$B$_{1-x}$C system can be described by
choosing the tight-binding parameters of the AC or BC system with probabilities
$x$ and $1-x$, respectively. We demonstrate this by exact diagonalization
of finite but large supercells and by means of calculations within the established
coherent potential approximation (CPA). We apply this treatment to the II-VI system
Cd$_x$Zn$_{1-x}$Se, to the III-V system In$_x$Ga$_{1-x}$As
and to the III-nitride system Ga$_x$Al$_{1-x}$N.
%
%}
\end{abstract}



% \titlerunning{Theory of band gap bowing of disordered substitutional II-VI 
% and III-V  alloys}
\maketitle

%==============================================================
% Introduction
%==============================================================
\section{Introduction}
Disordered semiconductor alloys of the type A$_x$B$_{1-x}$C can be experimentally
realized for many technologically important semiconductor materials AC and BC
(like GaAs and InAs, CdSe and ZnSe, GaN and AlN and many more)
by  substituting the cation A by the B counterpart.
For a wide class of materials,
the value  $x$ can be adjusted over a large part of or even
the whole concentration range, depending on the miscibility gap of the alloy system.
As the electronic and optical properties vary continuously with $x$,
the electronic properties (e.g. band gap and thus emission and absorption frequency,
dielectric constant, etc.) can be tailored by choosing the appropriate concentration.
Therefore, these semiconductor alloys have many applications, not only as bulk
semiconductors but also in low-dimensional structures like quantum wells,
quantum wires and quantum dots. 
%, which makes them particularly
%suitable in optoelectronic and especially photonic devices.

In this paper, we show that the electronic properties of such substitutional
semiconductor alloys of the type A$_x$B$_{1-x}$C can be described and understood
on the basis of empirical tight-binding models (ETBM)
for the pure semiconductors AC and BC.
Once the tight-binding (TB) parameters of AC and BC are known, 
the ETBM for the substitutionally disordered A$_x$B$_{1-x}$C alloy is obtained
by choosing the AC TB parameters with probability $x$ and the BC TB parameters with
probability $1-x$.
The electronic eigenenergies and thus the density of states and the band gap
of this disordered system can then be determined
by exact diagonalization of a finite (but large) supercell.
Unfortunately, it is necessary to consider a large number $N$ of microscopically
distinct configurations for each concentration $x$,
as well as sufficiently large supercells, consisting of several thousand primitive
unit cells at least.
As this procedure is numerically expensive,
we also considered and applied well established approximations
for the treatment of disordered systems,
namely the coherent potential approximation (CPA) and the 
virtual crystal approximation (VCA).
While the CPA is a self-consistent Green function approximation,
 the much simpler (but still frequently applied) VCA
corresponds to a static mean-field treatment of the disorder
by replacing the randomly fluctuating potential by an average potential.

We apply our treatment to the zincblende 
phase of three interesting materials,
namely to the II-VI semiconductor alloy system Cd$_x$Zn$_{1-x}$Se,
to the technically established III-V alloy In$_x$Ga$_{1-x}$As and to the III-nitride system Ga$_x$Al$_{1-x}$N.
In a previous paper [\onlinecite{mourad_band_2010}] it has already been shown
that the supercell exact diagonalization method yields excellent agreement with
experimental results obtained for Cd$_x$Zn$_{1-x}$Se.
But as [\onlinecite{mourad_band_2010}] was a joint paper together with experimentalists,
the theoretical method was not yet described and discussed in detail
and no comparison with  CPA and VCA results for this material has been presented there.
Whereas the VCA, in general, yields no band gap bowing,
the CPA may even overestimate the  bowing,
as we will show in the present paper.
We particularly present results obtained within combinations of CPA and VCA,
treating the site-diagonal disorder  in CPA and the off-diagonal disorder in VCA.
Furthermore, we explicitly discuss the influence of a properly
chosen basis set for the bowing results.

Also for In$_x$Ga$_{1-x}$As, reasonable results for the band gap bowing are obtained
in agreement with literature values as obtained by experiment. 
For  Ga$_x$Al$_{1-x}$N the band gap bowing is rather small
and the VCA results are not too bad at all;
a crossover from a direct band gap to an indirect band gap behavior
as a function of $x$ is obtained in all three applied approaches.

This paper is organized as follows: In Section \ref{sec:theory},
we present the necessary theoretical background for our calculations of the
electronic properties of A$_x$B$_{1-x}$C semiconductor bulk alloys. It contains
a brief description of the employed tight-binding model, the CPA and the modelling
on a finite ensemble of supercells. 
Section \ref{sec:results} contains a detailed analysis
of the results that we obtained for the electronic properties 
of Cd$_{0.5}$Zn$_{0.5}$Se with the
different models. Furthermore, we carefully analyze
the resulting band gap bowing 
of Cd$_x$Zn$_{1-x}$Se, In$_x$Ga$_{1-x}$As and  Ga$_x$Al$_{1-x}$N
 and discuss the applicability of
the CPA and VCA to these systems.

%==========================================================================
%                                  Theory
%==========================================================================
\section{Theory \label{sec:theory}}
%==========================================================================
\subsection{Empirical tight-binding model for pure bulk semiconductors:
\label{subsec:ETBMbulk}}
%==========================================================================
%

% %
% \begin{table}[t]%[H] add [H] placement to break table across pages
% \caption{Zero temperature material parameters for zincblende CdSe and
% ZnSe, taken from Refs.~[\onlinecite{bla,bla,bla}. Necessary modifications to some of these parameters, in order to account for different experimental conditions will be adressed in the text.}
% \label{tab:param}
% \begin{ruledtabular}
% \begin{tabular}{llcc}
% Parameter & Description &CdSe & ZnSe\\
% \hline
% $a$ (\AA) & lattice constant & 6.078 & 5.668\\
% $E_{\mathrm{g}}$ (eV) & band gap& 1.76 & 2.82\\
% $\Delta E_{\mathrm{vb}}$ (eV) & valence band offset & 0.22 & 0.00\\
% $X_1^\mathrm{c}$  (eV) & X-point energy of CB & 2.94 & 4.41 \\
% $X_5^\mathrm{v}$  (eV) & X-point energy of HH/LH band & -1.98 & -2.08  \\
% $X_3^\mathrm{v}$  (eV) & X-point energy of SO band & -4.28 & -5.03  \\
% $\Delta_{\mathrm{so}}$ (eV) & spin-orbit splitting & 0.41 & 0.43 \\
% $m_\mathrm{e}$ ($m_0$) & CB effective mass & 0.120 & 0.147 \\
% $\gamma_1$ & & 3.33 & 2.45  \\
% $\gamma_2$  & Kohn-Luttinger parameters & 1.11 & 0.61\\
% $\gamma_3$ & & 1.45 & 1.11  
% \end{tabular}
% \end{ruledtabular}
% \end{table}
% %

 %
\begin{figure*}
    \centering
    \includegraphics[width=0.95\linewidth]{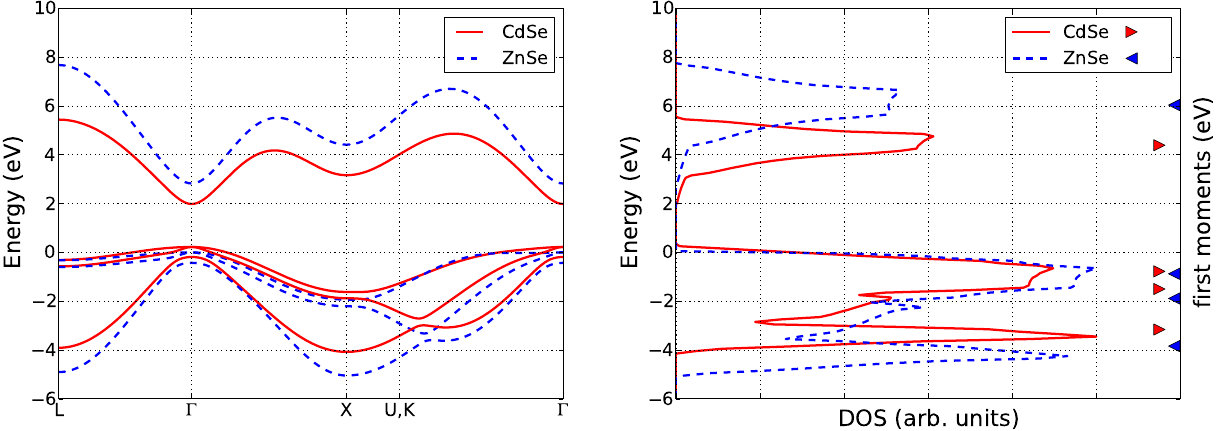}
    \caption{(Color online) Band structure (left) and DOS (right) of CdSe and ZnSe,
             calculated in the $sp^3$ EBOM.
             The energies of CdSe are already shifted by a valence band offset
             $\Delta E_\text{vb} = 0.22$ \eV with respect to ZnSe. For later use,
             the energetic position of the first moments of the bands is also
             given.   
             See text and appendix for details.}
    \label{fig:BS_DOS_CdSe_ZnSe}
\end{figure*}

We know from textbook solid state physics that one complete
basis set of the translationally
invariant pure bulk system is given by the Wannier states $\ket{n \f{R}}$, where $n$ is
the band index and $\f{R}$ denotes the unit cell where the state is predominantly localized.
The Wannier states are connected to the itinerant Bloch states $\ket{n \f{k}}$, where $\mathbf{k}$ is any wave vector in the first Brillouin zone (BZ), by means of a discrete Fourier transformation. 
The ETBM, as originally introduced by
Slater and Koster [\onlinecite{slater_simplified_1954}], uses a finite linear combination of atomic
orbitals on the basis sites as ansatz for the Wannier states.
The matrix elements are then fitted in order
to sufficiently reproduce prominent band structure features.

Because the atomic states, and hence the Wannier states, are not explicitly used in an
ETBM, one has the additional freedom to directly assume a finite basis set of localized
Wannier-like states. As the Wannier basis carries the translational invariance of the
crystal, these states merely have to be assigned to the sites $\f{R}$ of the underlying
Bravais lattice.
This parametrization scheme is also known as effective bond-orbital model (EBOM) in the
literature.

In this paper, we will use a $sp^3$ basis per spin direction $\left\{\uparrow,
\downarrow\right\}$ to reproduce
$N_\alpha=8$ bands for crystals with zincblende structure: 
One spin-degenerate $s$-like conduction band (CB)
and three spin-degenerate $p$-like valence bands (VB), in detail the heavy hole (HH),
light hole (LH) and split-off (SO) band. Thus, $\f{R}$ labels the $N_\f{R}$ sites of
the underlying fcc Bravais lattice, and
the TB matrix elements of the bulk Hamiltonian $H^{\text{bulk}}$  are
\begin{equation}\label{eq:EBOMme}
    E_{\alpha \alpha'}^{\mathbf{R} \mathbf{R'}} = 
                              \matrixelem{\f{R} \alpha}{H^{\text{bulk}}}{\f{R}' \alpha'},
\end{equation}
with the orbitals
\begin{equation}\label{eq:sp3basis}
    \left| \mathbf{R} \alpha \right\rangle, \quad \alpha \in 
    \left\lbrace s\uparrow,p_x\uparrow,p_y\uparrow,p_z\uparrow,
    s\downarrow,p_x\downarrow,p_y\downarrow,p_z\downarrow \right\rbrace.
\end{equation}
The translational invariance in the pure crystal reduces the
$N_{\alpha}N_\f{R} \times N_{\alpha}N_\f{R}$ matrix problem to a
$N_{\alpha}\times N_{\alpha}$  problem for each $\f{k}$ [\onlinecite{slater_simplified_1954}], 
which will not be the case for disordered
systems and in the later introduced supercell method.
The band structure $E(\mathbf{k})$ is then given by the solution of the eigenproblem 
\begin{equation}
    \sum_{\alpha'} \, \sum_{\mathbf{R}} e^{i \mathbf{k} \cdot \mathbf{R} }
    E_{\alpha  \alpha'}^{\mathbf{0} \mathbf{R}} \, c_{\alpha'}(\mathbf{k})
    = E(\mathbf{k}) \, c_{\alpha}(\mathbf{k}).
\label{eq:eigenproblem}
\end{equation}
If one restricts the non-vanishing matrix elements, Eq.\,(\ref{eq:EBOMme}), up to a
finite neighborhood 
and makes proper use of symmetry relations, the TB matrix elements can be expressed as
a function of a set of material parameters  of either the AC or BC material (e.\,g.\,
CdSe and ZnSe). We will use the parametrization scheme of Loehr [\onlinecite{loehr_improved_1994}],
which includes coupling up to second nearest neighbors.
The parameters are given in the appendix.  Figure \ref{fig:BS_DOS_CdSe_ZnSe} shows
the band structure and density of states (DOS) as calculated in the $sp^3$ EBOM 
for cubic CdSe and ZnSe, respectively. 

Despite the small basis set, this TB model notably allows
for the reproduction of a realistic bandstructure and bandwidth throughout the whole BZ.
This is a typical feature of parametrizations with Wannier-like basis states
[\onlinecite{mourad_multiband_2010}] and makes the EBOM especially 
suitable for the purpose of the present work,
 as both the CPA and the supercell calculations
for the A$_x$B$_{1-x}$C alloys  will require a realistic input
DOS for the pure
cases $x=0$ and $x=1$ as starting point---the relative position of the band centers
and the bandwidths will crucially influence the alloy properties.
 We have also already mentioned
in the introduction that the inclusion
of further bands would significantly increase the computational effort.
Finally, the resolution on the scale of unit cells instead of atomic
sites will turn out as
favorable for the simulation of the ternary A$_x$B$_{1-x}$C
 materials, as it will allow for an unambiguous
assignment of the alloy lattice sites to either AC or BC (see Sec.\,
\ref{subsec:supercellTBcalculation}).

%==========================================================================
\subsection{The coherent potential approximation
\label{subsec:CPA}}
%==========================================================================

The basic principle of the CPA has independently been developed by several
groups in the late sixties of the last century
[\onlinecite{soven_coherent-potential_1967}][\onlinecite{taylor_vibrational_1967}][\onlinecite{onodera_persistence_1968}].
The best known formulation is certainly 
the work of Soven [\onlinecite{soven_coherent-potential_1967}], which explicitely dealt with
the calculation of the electronic DOS of substitutionally disordered one-dimensional systems.
An excellect and comprehensive general introduction into the CPA formalism  
can be found in [\onlinecite{matsubara_structure_1982}].

The simplest form of the CPA for an alloy assumes uncorrelated substitutional
disorder of the respective species. If all sites are indistinguishable
(i.\,e.\, no division into sublattices is necessary), as it is the case in the 
spatial discretization in the EBOM,
the probability of finding either species is just given by the concentrations $x$ and
$1-x$.

For the sake of clarity, we will supress the band  and orbital indices until further notice.
In order to be consistent with the common CPA literature, let
us denote the site diagonal ($\f{R}=\f{0}$) TB matrix elements of the AC or BC material
as $v_\text{AC/BC}$:
\begin{displaymath}
    v_\text{AC/BC} =  E_{\text{AC/BC}}^{\mathbf{0} \mathbf{0}} =
    \matrixelem{\f{0}}{H^{\text{bulk}}_{\text{AC/BC}}}{\f{0}}.
\end{displaymath}
We will now briefly specify the most important assumptions that enter the CPA
(see e.\,g.\,[\onlinecite{matsubara_structure_1982}] for details of the derivation):
\begin{enumerate}
    \item
    The disorder is confined to the diagonal elements. The off-diagonal
    (or \textit{hopping}) matrix elements with $\f{R} \neq \f{R}'$ are either identical
    for both species or can approximately be replaced by a common value, e.\,g.\,
    the VCA average. If we now separate the TB Hamiltonian of the alloy 
    $H$ into a diagonal part $V$
    and an off-diagonal part $W$, such that $H=V+W$,
    only $V =  \sum_{\f{R}} v^{\f{R}} \ket{\f{R}}\bra{\f{R}}$ is site-dependent, as
    $v^\f{R} = v_\text{AC/BC}$, depending on the species on the site $\f{R}$.
    The operator $W$ is still translationally invariant under translations
    by $\f{R}$
    and therefore remains diagonal with respect to  Bloch states $\ket{\f{k}}$.
    \item 
    The configurational average $\li \ldots \re$ over the resolvent of
    $H$ defines an effective Hamiltonian $H_\text{eff}(z)$:
    \begin{eqnarray}
        \li (z \mathds{1} - H)^{-1} \re &\equiv& \left[z \mathds{1} - H_\text{eff}
        (z)\right]^{-1} \nonumber\\ 
                  &\equiv&  \left[z \mathds{1} - \Sigma (z) - W\right]^{-1}.
                  \label{eq:Heff}
    \end{eqnarray}
    Here, $\mathds{1}$ is the identity operator and $z$ is in the complex energy plane,
    containing the energy axis
    $E = \real{(z)}$. The self-energy operator
    $\Sigma(z) = H_\text{eff}(z) - W$ absorbs
    the influence of the disorder on the microscopic scale. Note that $H_\text{eff}$ will
    in general be non-hermitian.
    \item
    Due to the single-site nature of the CPA, the self-energy is diagonal
    in every representation. Furthermore, its matrix elements $\Sigma^0(z)$ are neither
    dependent on $\f{k}$ nor on $\f{R}$.
\end{enumerate}
In order to obtain the self energy matrix elements $\Sigma^0(z)$, which uniquely 
define the effective medium, we have to solve the CPA equation for two constituents:
\begin{equation}
    \frac{x \left[ v_{\text{AC}} - \Sigma^0(z) \right] }
    {1- \left[ v_{\text{AC}} - \Sigma^0(z)  \right] G_{\f{R}}(z)} 
  + \frac{(1-x)  \left[ v_{\text{BC}} - \Sigma^0(z) \right] }
    {1- \left[ v_{\text{BC}} - \Sigma^0(z)  \right] G_{\f{R}}(z)}= 0.
\label{eq:CPAbinary}
\end{equation}
The complex-valued $G_{\f{R}}(z)$ is the configurationally averaged 
one-particle Green function in Wannier representation (more precisely its $\f{R}$-diagonal
element).
Although $G_\f{R}(z)$ does not depend on $\f{R}$ in the CPA, we will keep the index to
clearly distinguish it from its Bloch representation $G_{\f{k}} (z)$.
The two representations are connected by
\begin{eqnarray}
G_{\f{R}}(z) &=& %\phantom{\frac{1}{N}}
            \matrixelem{\f{R}}{\li (z \mathds{1} - H)^{-1} \re}{\f{R}} \nonumber \\
    &=& \frac{1}{N_\f{k}} \sum\limits_{\f{k}} \matrixelem{\f{k}}{ \li (z \mathds{1} - H)^{-1}
        \re}{ \f{k}} \nonumber\\
    &\stackrel{\text{Eq.}\,(\ref{eq:Heff})}=& \frac{1}{N_\f{k}} \sum\limits_{\f{k}} \left[z - \Sigma^0(z) -
            \matrixelem{\f{k}}{W}{  \f{k}}  \right]^{-1} \nonumber \\
    &\equiv &  \frac{1}{N_\f{k}} \sum\limits_{\f{k}} G_{\f{k}} (z),
\label{eq:GBloch}
\end{eqnarray}
with $N_\f{k}$ as the number of wave vectors in the first BZ or the corresponding
irreducible wedge (as $N_\f{k}=N_\f{R}$ and $G_\f{R} = G_{\f{R}+\f{R'}}$, this relation
follows directly from the invariance of the trace under unitary transformations).

Under certain conditions, the summation over all $\f{k}$-values can be avoided by the
introduction of a sufficiently smooth 
 DOS $g_W$ for the off-diagonal part $W$:
\begin{equation}
g_{\text{W}}(E) = \frac{1}{N_{\f{k}}} \sum\limits_{\f{k}} \delta(E - \matrixelem{\f{k}}{W}{\f{k}}).
\end{equation}
Now the calculation of $G_{\f{R}}(z)$ can be performed as a one-dimensional
integration/summation over the energy axis:
\begin{eqnarray}
    G_{\f{R}}(z) &=&\frac{1}{N_\f{k}} \sum\limits_{\f{k}} \frac{1}{z - \Sigma^0(z)
                                            - \matrixelem{\f{k}}{W}{\f{k}}} \nonumber \\
%     &\approx& \sum\limits_{i}g_{W}(E_i) \Delta E_i \frac{1}{z-\Sigma^0(z) - E_i}\nonumber \\
    &=& \int dE   \frac{g_{W}(E)}{z-\Sigma^0(z) - E}\nonumber \\
    &\approx&  \sum\limits_{i} g_{W}(E_i) \int\limits^{E_{i+1}}_{E_i} dE 
                     \frac{1}{z-\Sigma^0(z) - E} \nonumber\\
    &=& \sum\limits_{i} g_{W}(E_i) \left\{ \ln \left[z-\Sigma^0(z)-E_i\right] \right.
         \label{eq:GWannier}\\
     & &\hspace{2.0cm}- \left. \ln \left[z-\Sigma^0(z)-E_{i+1} \right] \right\}. \nonumber         
\end{eqnarray}
The approximation holds if $g_W(E)$ is approximately constant on
the interval $[E_i,E_{i+1}]$. 

The equations
(\ref{eq:CPAbinary}) and optionally (\ref{eq:GBloch}) or (\ref{eq:GWannier}) can be
 solved in a self-consisted manner
in order to obtain $\Sigma^0(z)$ and $G_{\f{R}}(z)$. We can then directly obtain 
further quantities for the effective medium; e.\,g.\,the configurationally
averaged DOS $g(E)$ per lattice site is given as
\begin{eqnarray}
    g(E) &=& - \frac{1}{\pi} \lim\limits_{z \rightarrow E^+} \imag \frac{1}{N_{\f{R}}}
               \trace\left\lbrace \left[z \mathds{1} - H_\text{eff}(z)\right]^{-1}
               \right\rbrace \nonumber\\ 
         &=& -\frac{1}{\pi} \lim\limits_{\delta \searrow 0} \imag 
                       G_{\f{R}}(E + i \delta).
               \label{eq:ensembleDOS}
\end{eqnarray}

As a rule of thumb, the CPA is known to yield very good results in two limit cases 
[\onlinecite{matsubara_structure_1982}]:
\begin{enumerate}
    \item
    The \emph{weak scattering limit}, where the difference of the 
    first moments of the substituents (aka the centers of gravity of the bands)
    is smaller than the respective bandwidths.
    \item The \emph{split band limit} or \emph{atomic limit}, where the substituents'
    first moments are either
    sufficiently far apart or the bandwidths are
    small, so that the respective bands do not overlap.
\end{enumerate}
Additionally, it of course also gives the correct pure limit for $x\rightarrow0$
and $x\rightarrow 1$. Also, in the limit case of vanishing difference of the moments,
the CPA reduces to the VCA.
%==========================================================================
\subsection{Multiband CPA + ETBM
\label{subsec:CPAETBM}}
%==========================================================================

The formal extension of the CPA to multiband TB models is straightforward and has been
used in different levels of detail to qualitatively examine the electronic properties of
disordered alloys (see e.\,g.\,[\onlinecite{stroud_band_1970}][\onlinecite{krishnamurthy_band_1986}]
for  Si$_x$Ge$_{1-x}$, [\onlinecite{faulkner_electronic_1976}] for Pd$_x$H$_{1-x}$,
[\onlinecite{hass_electronic_1983}] for Cd$_x$Hg$_{1-x}$Te
or [\onlinecite{laufer_tight-binding_1987}] for palladium-noble-metal alloys).
%

%=========================================
\subsubsection{$sp^3$ representation}

In the $sp^3$ EBOM, the localized basis is now given by the TB 
orbitals $\ket{\f{R} \alpha}$.
Hence, Eq.\,(\ref{eq:CPAbinary}) has to be replaced by the corresponding
matrix equation, with
\begin{eqnarray}
  v_{\text{AC/BC}} &\rightarrow&  \mtrx{v}_{\text{AC/BC}}^{sp^3}
             \equiv \left[ E_{\alpha \alpha',\text{AC/BC}}^{\mathbf{0} \mathbf{0}} \right],  \nonumber\\
  \Sigma^0(z) &\rightarrow& \mtrx{\Sigma}^{sp^3}(z)
         \equiv \left[ \Sigma^{\alpha \alpha'}(z) \right],\\
  G_{\f{R}}(z) &\rightarrow& \mtrx{G}_{\f{R}}^{sp^3}(z)
      \equiv \left[ G^{\alpha \alpha'}_{\f{R}}(z) \right] \nonumber
\label{eq:matrixsubstitutions}
\end{eqnarray}
as $4 \times 4$ matrices per spin direction (here, the square brackets denote matrices). %
The effective Hamiltonian matrix $\mtrx{H}^{sp^3}_\text{eff}(z)$ 
in the $sp^3$ TB scheme can be obtained by the substitutions
\begin{eqnarray}
 E_{\alpha \alpha'}^{\mathbf{0} \mathbf{0}} &\stackrel{\text{CPA}}\longrightarrow&
                     \Sigma^{\alpha \alpha'}(z), 
                     \label{eq:sp3selfenergy}\\
 E_{\alpha \alpha'}^{\mathbf{R} \mathbf{R'}} &\stackrel{\text{VCA}}\longrightarrow& 
                   x \, E_{\alpha \alpha', \text{AC}}^{\mathbf{R} \mathbf{R'}}
                  + (1-x) \, E_{\alpha \alpha', \text{BC}}^{\mathbf{R} \mathbf{R'}},
\end{eqnarray}
i.\,e.\, the hopping matrix elements are approximated in the VCA. Like in 
Eq.\,(\ref{eq:GBloch}), we then obtain the Green function via BZ summation,
\begin{equation}
    \mtrx{G}_{\f{R}}^{sp^3}(z) = \frac{1}{N_\f{k}} \sum\limits_{\f{k}}
    \left[z \mtrx{1} -\mtrx{H}^{sp^3}_\text{eff}(z)\right]^{-1} ,    
\label{eq:CPAGreensp3}
\end{equation}
where $\mtrx{1}$ is the $8 \times 8$ identity matrix.
The matrix $\mtrx{G}_{\f{R}}^{sp^3}(z)$ 
has non-vanishing off-diagonal elements $G^{\alpha \alpha'}_{\f{R}}(z)$,
as the TB Hamiltonian
is not diagonal in the orbital basis $\ket{\f{R} \alpha}$. 

%======================================================
\subsubsection{Band-diagonal (Wannier) representation\label{subsubsec:WannierRep}}

Alternatively, we can again calculate Green's function via energy integration. In a
band-diagonal (Wannier-like) TB representation,
the site-diagonal TB matrix element \\
$E^{\f{0}}_{n} = \matrixelem{\f{0} n}{H^\text{bulk}}{\f{0} n}$
of the $n$-th band can be calculated from the first moment
\begin{equation}
 E^{\f{0}}_{n} = \int dE E \, g^n(E),
\end{equation}
with $g^n(E)$ as 
the DOS of band $n$ (normalized to unity). Therefore, we 
can assign the moments of the AC or BC system to a corresponding self-energy:
\begin{equation}
E^{\f{0},\text{AC/BC}}_n  \rightarrow \Sigma^{n}(z).
\label{eq:Wannierselfenergy}
\end{equation}
The analogon of equation (\ref{eq:GWannier}) is then the
band-diagonal Green function matrix $\mtrx{G}_{\f{R}}^{\text{wn}}(z)$
(denoted by ``wn'' for ``Wannier'' from now on) with elements
\begin{eqnarray}
   G^{nn'}_{\f{R}}(z) &\approx& \sum\limits_{i} g^n_{\text{W}}(E_{n,i})
    \left\{ \ln \left[ z-\Sigma^n(z)-E_{n,i}\right] \right. \nonumber \\
    & & \hspace{1cm}  - \left. \ln \left[z-\Sigma^n(z)-E_{n,i+1}\right] \right\} \delta_{nn'},
\label{eq:CPAGreenbanddiagonal}
\end{eqnarray}
with $g^n_W(E) = g^{\text{VCA}}(E-E^\f{0}_n)$ as the VCA DOS of the $n$-th band,
shifted by $E^\f{0}_n$.
This procedure allows us to calculate the DOS of the off-diagonal operator $W$ in the
Wannier basis by diagonalization of the EBOM Hamiltonian in the VCA. Notably, we do
not have to explicitly know the corresponding matrix elements
$\matrixelem{\f{R} n}{W}{\f{R'} n}$.

%===================================================
\subsubsection{Calculation of electronic properties}

\begin{figure*}
    \centering
    \includegraphics[width=0.95\linewidth]{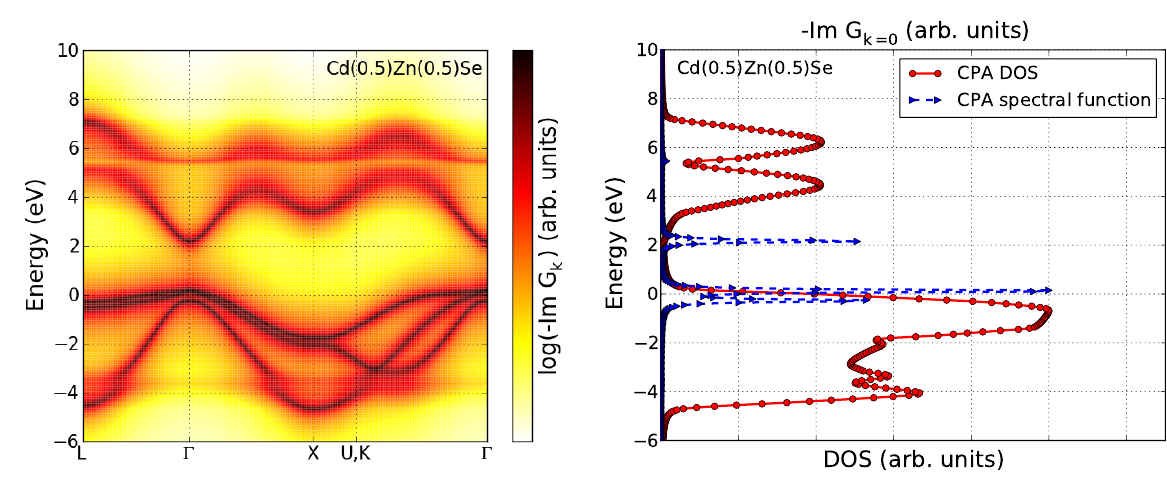}
    \caption{(Color online) Complex band structure (left), DOS and 
             spectral function at $\f{k}=\f{0}$
             (right) for Cd$_{0.5}$Zn$_{0.5}$Se, calculated in the combination of 
             the CPA  and the $sp^3$ EBOM. To enhance contrast,
             we have color-coded the logarithm of $-\imag G_\f{k}$.
             The spectral function and the DOS are normalized to a common scale.}
    \label{fig:CPA_BS_DOS_CdZnSe}
\end{figure*}

Finally, the CPA + ETBM DOS of the alloy can be calculated
by either tracing over the orbital index (in the $sp^3$ basis)
or the band index (in the Wannier basis)
of the matrix elements:
\begin{eqnarray}
g^{sp^3}(E) &=&  -\frac{1}{\pi}\lim\limits_{\delta \searrow 0} \imag \sum\limits_\alpha  G^{\alpha \alpha}_{\f{R}}(E + i \delta), \\
g^{\text{wn}}(E) &=&  -\frac{1}{\pi} \lim\limits_{\delta \searrow 0}\imag \sum\limits_n  G^{n n}_{\f{R}}(E + i \delta)   .
\label{eq:ensembleDOS2}
\end{eqnarray}
Although the trace of a matrix is of course independent of the representation used,
it should be noted that $g^{sp^3}$ and $g^{\text{wn}}$ are not identical:
The self-energies, which define the effective medium, replace different quantities
[see Eqs.\,(\ref{eq:sp3selfenergy}) and (\ref{eq:Wannierselfenergy})], which
results in different operators $H_{\text{eff}}$. 

Due to the translational invariance of the self energy matrix elements and thus 
$H_{\text{eff}}$, we can define a complex band structure of
the medium. The electronic excitations of the effective medium can be assigned 
to quasiparticles with a modified 
dispersion relation
\begin{equation}  
    E^{\text{Q}}_n(\f{k}) = \matrixelem{\f{k}n}{W}{\f{k}n} + \real \Sigma^n(E^{\text{Q}}_n). 
    \label{eq:quasiparticleBS}
\end{equation} 
The corresponding CPA one-particle spectral function
\begin{equation}
S^n_{\f{k}}(E) = -\frac{1}{\pi} \lim\limits_{\delta \searrow 0}
                   \imag G^{nn}_{\f{k}}(E + i\delta),
\label{eq:CPAspectralfunction}
\end{equation}
will in general be broadened in case of disorder $(0 < x < 1)$, as the non-vanishing
imaginary part of $\Sigma^n$ accounts for a finite lifetime
$\tau \propto 1/ \imag \Sigma^n(E)$ of the excitation.

As an example, Fig.\,\ref{fig:CPA_BS_DOS_CdZnSe} visualizes the complex band
structure $E^{\text{Q}}_n(\f{k})$ of Cd$_{0.5}$Zn$_{0.5}$Se on the left,
calculated in the combination of the CPA and the $sp^3$ EBOM. On this
energy scale over the whole bandwidth, the results from the Wannier
and the $sp^3$ representation are not distinguishable. The color-coding
clearly shows the $\f{k}$-dependent broadening of the band structure due to 
finite lifetime effects.
The figure on the right additionally shows the corresponding CPA DOS and the spectral
function at the BZ center  for all four  spin-degenerate bands, i.\,e.\,
\begin{displaymath} 
S_{\f{k}=\f{0}}(E) = -\frac{1}{\pi} \lim_{\delta \searrow 0}\imag 
\sum_n G^{nn}_{\f{k}=\f{0}}(E + i\delta).
\end{displaymath}
Note that we have chosen a relatively large imaginary part 
$\delta=0.05$ \eV for the energy axis,
which leads to a small but non-vanishing DOS
in the band gap region. We furthermore
used $\#\f{k}\approx 10^6$ values in the irreducible 
BZ of the fcc lattice. The peak structure of the $\f{k}=\f{0}$
quasiparticle excitations at the edges of the band gap region is clearly visible.
A qualitative and quantitative discussion of the electronic structure in comparison
with VCA and supercell results will follow in Section \ref{subsec:overallComparison}.

%==========================================================================
\subsection{Supercell tight-binding calculation in the EBOM
\label{subsec:supercellTBcalculation}}
%==========================================================================

% In this section, we will briefly introduce the calculation of the electronic
% properties of the A$_x$B$_{1-x}$C alloy in the framework of tight-binding 
% calculations on finite supercells. For a detailed 

If the Hamilton operator $H$ is no longer
translationally invariant, the TB approach is not reducable
to the form (\ref{eq:eigenproblem}) and we are left with the
$N_{\alpha}N_\f{R} \times N_{\alpha}N_\f{R}$ matrix eigenvalue equation
\begin{equation}
    \sum_{\alpha'\f{R'}}
    \matrixelem{\f{R}\alpha}{H}{\f{R'}\alpha'}  c_{\alpha'\f{R'}}
    = E \, c_{\alpha\mathbf{R}}.
\label{eq:SCeigenproblem}
\end{equation}
We will now use a finite supercell with periodic boundary conditions. Like in 
calculations for zero-dimensional nanostructures
[\onlinecite{marquardt_comparison_2008}][\onlinecite{schulz_multiband_2009}][\onlinecite{mourad_multiband_2010}][\onlinecite{mourad_multiband_2010-1}], we will furthermore assume that
\begin{equation}
    \matrixelem{\f{R}\alpha}{H}{\f{R'}\alpha'} 
         \approx \matrixelem{\f{R}\alpha}{H^\text{bulk}}{\f{R'}\alpha'}
          =  E_{\alpha  \alpha'}^{\mathbf{R} \mathbf{R'}}. 
\label{eq:}
\end{equation}
Again assuming uncorrelated substitutional disorder in the  A$_x$B$_{1-x}$C alloy,
each primitive cell will be occupied by the AC or BC basis, where the
probability of finding either an AC or BC pair in the unit cell at $\f{R}$ is
directly given by the concentrations $x$ and $1-x$.
Building the Hamilton matrix for the supercell, we will therefore
use the  matrix elements of the pure
AC or BC material for the corresponding lattice sites. The valence band offset 
between the two materials is incorporated by shifting the respective site-diagonal
matrix elements by a value $\Delta E_{\text{vb}}$. Hopping matrix elements
between unit cells of different material are approximated by the arithmetic average
of the corresponding AC or BC values. Although this approach for the incorporation
of the energy offset and the hopping between two materials is very simple, it gives 
the correct limit in the pure case. In the phase separation case (a limit of which the
propability is practically zero in case of uncorrelated disorder),
it would furthermore lead to an interface treatment that has been extensively 
tested in calculations for low-dimensional heterostructures 
[\onlinecite{marquardt_comparison_2008}][\onlinecite{schulz_multiband_2009}][\onlinecite{mourad_multiband_2010}].
Specifically, the results turn out to be insensitive to small variations
of the hopping between two materials (e.\,g.\, the usage of a geometric
instead of an arithmetic average).

The usage of the EBOM, i.\,e.\,the usage of Wannier-like orbitals situated
at the sites of the Bravais lattice in the supercell approach, has several advantages
over similar approaches which use an ETBM with discretization on atomic sites:
\begin{enumerate}
    \item
    Even when introducing disorder on a microscopic scale,
    each lattice site can be unambigously assigned to one site-diagonal TB matrix
    element and the corresponding band structure parametrization for either AC or BC.
    In ETBM supercell calculations with atomic resolution, each anion of the type C 
    will locally be surrounded by a different number of A or B cations, thus making an
    assignment of the diagonal elements of the anions to the band structure of either
    AC or BC impossible.
    
    It is common to then use either a concentrationally averaged 
    VCA value or to determine this matrix element as a weighted average of the
    C matrix elements for AC and BC, depending upon the number of nearest-neighbour
    atoms A or B [\onlinecite{tit_absence_2009}][\onlinecite{boykin_approximate_2007}].
    This will lead to an effectively more coarse-grained resolution as in the case of
    the EBOM, as the latter model only has to average the intersite hopping matrix
    elements (which typically differ on a scale of $10$ \meV in materials with
    moderate lattice mismatch).
    In the Cd$_{x}$Zn$_{1-x}$Se system for example,
    the site diagonal matrix elements for the Se anions in CdSe
    and ZnSe (see e.\,g.\,Refs. [\onlinecite{tit_absence_2009}]
    and [\onlinecite{schulz_tight-binding_2005}]) differ to a larger extent than the EBOM 
    hopping matrix elements when using a congruent set of input parameters.
    \item
    All influences which originate from effects on a smaller length scale, like the
    difference in the AC and BC bond lengths, are absorbed into the values of the
    corresponding TB matrix elements between the effective orbitals.    
    \item
    As the results are not sensitive
    to the exact treatment of the hopping between AC and BC sites,
    further effects that basically result in minor variations
    of the hopping matrix elements (like small bond angle changes due to relaxation)
    can be neglected in a first approximation.
\end{enumerate}

The numerical diagonalization (e.\,g.\,using standard numerical 
libraries like ARPACK/PARPACK) of the corresponding Hamiltonian for a fixed
concentration and a finite number $N$ of microscopically
distinct configurations gives the density of states (DOS) of the finite ensemble.
In order to obtain a meaningful DOS from the supercell calculations, we must first
appropriately define it.
Strictly speaking,  a macroscopic
alloy crystal represents just one realization; by dividing it into small portions,
we can nevertheless get subsystems that differ from each other on a microscopic
scale. In this sense, the DOS for one fixed concentration $x$ is then obtained
by the  average of the DOS for each finite ensemble.

To eliminate the influence of finite size effects, the number of lattice sites
$N_\f{R}$ as well as the ensemble size $N$ must be sufficiently large. We point to
previous work [\onlinecite{mourad_band_2010}] for a careful analysis of the convergency 
behaviour and will use the
recommendations throughout this paper. In a nutshell, a resolution of the
band edges up to $0.01$ \eV, which is the typical input accuracy for 
the material parameters, will require
supercells with $N_\f{R} \approx 2000$--$4000$ lattice sites and  $N\approx5$0 
microscopically distinct configurations per concentration. For a
discussion of properties on a larger energy scale, e.\,g.\, a comparison
of the DOS over the whole bandwidth, smaller supercells and ensemble 
sizes can be chosen.

In contrast to the CPA, the supercell approach can easily be augmented to 
simulate effects not only of \emph{configurational}, but also of \emph{concentrational}
disorder. This can easily be achieved when we drop the constraint that the overall
concentration of AC sites  $N_\text{AC}^i/ N_\f{R}^i$ per configuration $i$ should equal
the point probability $x$ and occupy each lattice site independently.
In the limit of large $N$, we will of course have
$\lim_{N \rightarrow \infty} \sum_{i=1}^{N} N_\text{AC}^i/ N_\f{R}^i = x$, so that
the constraint is fulfilled  for sufficiently large ensemble numbers.
For most cases, this is closer to experimental reality anyway, as concentration
values are commonly averages over macroscopic volumes, e.\,g.\, by means of X-ray
diffraction [\onlinecite{mourad_band_2010}]. It also allows us to perform calculations for concentration values $x$
where $N_\f{R}/x \notin \mathds{N}$.

%==========================================================================
%                                  Results
%==========================================================================
\section{Results \label{sec:results}}

In this section, we will apply the CPA EBOM and the supercell EBOM 
to  cubic Cd$_{x}$Zn$_{1-x}$Se, In$_{x}$Ga$_{1-x}$As and Ga$_{x}$Al$_{1-x}$N.
In addition, we will
also add results that are obtained by a pure
VCA calculation.

Besides the fact that we have already shown the reliability of the supercell
EBOM for cubic Cd$_{x}$Zn$_{1-x}$Se in comparison with experimental results in 
[\onlinecite{mourad_band_2010}] (albeit for slightly different material parameters
to meet the experimental boundary conditions), this material systems
is also especially interesting
for a quantitative and qualitative analysis of the applicability of the CPA.
A closer look at the DOS and the energetic position of the first moments in
Fig.\,\ref{fig:BS_DOS_CdSe_ZnSe} reveals that the conduction bands of CdSe and ZnSe
neither fulfill the weak scattering nor the split band condition very well,
as the energetic range of the overlap is comparable to the difference 
of the first moments.
The In$_{x}$Ga$_{1-x}$As system under consideration will in contrast be closer
to the weak scattering limit.
This condition will also apply to the zincblende  Ga$_{x}$Al$_{1-x}$N alloy,
which additionally comprises a direct-indirect band gap transition 
at a certain mixing ratio.
 
%==========================================================================
\subsection{Comparison of the overall density of states of Cd$_{0.5}$Zn$_{0.5}$Se
\label{subsec:overallComparison}}
%==========================================================================

%
\begin{figure}
    \centering
    \includegraphics[width=0.95\linewidth]{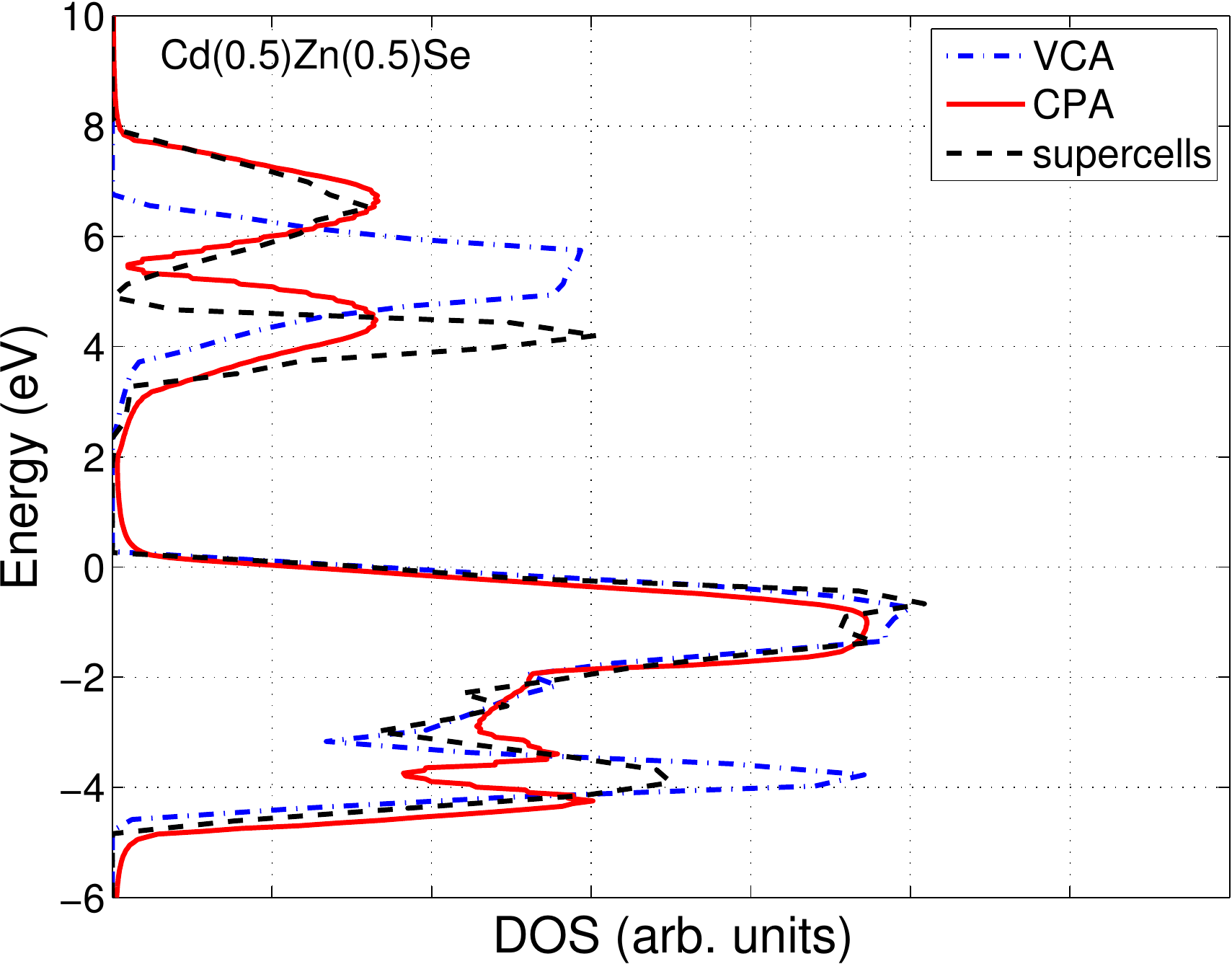}
    \caption{(Color online) Direct comparison of the DOS of
              Cd$_{0.5}$Zn$_{0.5}$Se, calculated in the combination
              of the VCA (blue), CPA (red) and supercell approach (black) and
              the EBOM. See  text for further details.}
    \label{fig:comparison_DOS}
\end{figure}

In this section, we will compare the overall DOS of
the A$_{x}$B$_{1-x}$C alloy as calculated with the CPA and the
supercell EBOM, using the example of Cd$_{0.5}$Zn$_{0.5}$Se,
along with results from the simple VCA.
The supercell calculations were performed with 20 microscopically
distinct configurations and on cubic supercells with 2048 lattice sites, i.\,e.\,
4096 atoms; the numerical parameters for the CPA calculation match those of Fig.\,
\ref{fig:CPA_BS_DOS_CdZnSe}. 

Overall, the alloy DOS of the valence bands is very similar in all three models.
This is not very surprising, as the substitutional disorder is 
restricted to the cations of the material, and the valence bands mainly stem from
atomic $p$-orbitals of the Se anions. However, the conduction band DOS
accordingly shows different features in the three models. 

It is clearly visible that the VCA DOS is an interpolation of the DOS of the pure 
CdSe and ZnSe material as depicted in Fig.\,\ref{fig:BS_DOS_CdSe_ZnSe}; aside
from an energetic shift of the bands, no new features arise for the alloy material.
   
Contrary to the VCA, the CPA gives a DOS with qualitative and 
quantitative features that exceed the results of simple
interpolation schemes by far. The conduction band visibly splits into two subbands.
When the resolution is further increased, a quasi-gap can be identified. 
The relative spectral weight of the two subbands is exactly given by 
the concentration ratio of $x/(1-x)=0.5/0.5=1$ of the substituents (this also holds
for all other values of $x$ with identifiable subband splittings).

Like in the CPA, the supercell EBOM also gives an alloy DOS of which the structure
is more complicated than the DOS of the constituents. The conduction band
DOS again splits into a two-subband structure, where the relative spectral
weight equals the concentration ratio of the substituents.
Furthermore, the two subbands
can now clearly be distinguished by a difference in their shape. Additionally,
the quasi-gap is located at a lower energy than in the CPA.

Overall, the fact that VCA totally fails to reproduce the additional features
of the alloy's conduction band was to be expected; the
one-electron potential which enters the hopping matrix elements
is not a self-averaging quantity (in the sense that
it can be replaced by its ensemble average for
a sufficiently large sample), as opposed to the one-electron
Green's function [\onlinecite{kohn_quantum_1957}].

The artificial symmetry in the conduction subband structure in the CPA
is an artefact of the usage of concentrationally dependent,
but nevertheless common intersite hoppings. This mean-field approach for the
hopping matrix elements directly carries over to the shape of the DOS (this can
most easily be seen in the Wannier basis,
as the shape and the bandwidth is eventually determined by the hopping
elements, while the band moments are given by the site-diagonal elements,
see Sec.\,\ref{subsubsec:WannierRep}).

%==========================================================================
\subsection{Comparison of the band gap bowing of Cd$_{x}$Zn$_{1-x}$Se
\label{subsec:comparisonBowing}}
%==========================================================================

%
\begin{figure}
    \centering
    \includegraphics[width=0.95\linewidth]{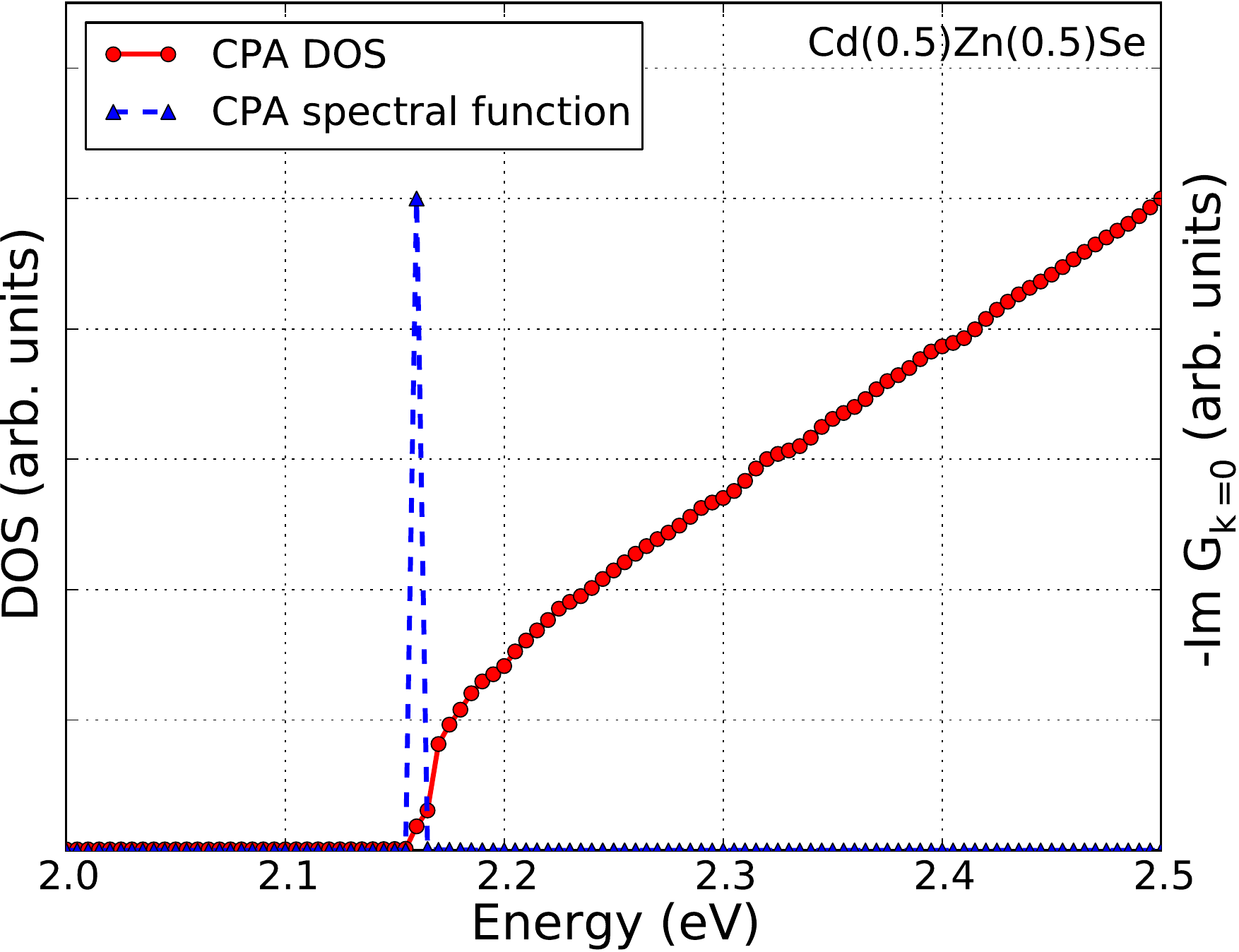}
    \caption{(Color online) Conduction band edge of Cd$_x$Zn$_{1-x}$Se, calculated in the 
            CPA. The peak of the spectral function at the BZ center coincides with the
            band edge within the energy resolution.
            See text for further details.}
    \label{fig:CPAbandedge}
\end{figure}

We will now extensively examine the accuracy of the CPA approach for properties on a 
smaller energy scale and use the example of the single-particle band gap for a 
qualitative as well as quantitative analysis.

Most alloyed bulk semiconductors show a more or less pronounced bowing of the band gap
$E_{\text{g}}$  as a function of the concentration $x$.
The simplest way to describe the
deviation from a linear behaviour is the assumption of a parabolic $E_{\text{g}}(x)$ curve
and therefore the use of a single, concentration-independent bowing parameter $b$, such
that
\begin{equation} \label{eq:bandgapbowing}
 E_{\text{g}}(x) = x \, E_{\text{g}}^{\text{AC}} + (1-x) \, E_{\text{g}}^{\text{BC}}
                    - x \, (1-x) \, b.
\end{equation}
Here, the indices AC and BC assign the properties of the pure binary materials.
In general, the literature values for $b$ show a surprisingly large variety even for
apparently comparable experimental conditions (the reader may check comprehensive review
articles like [\onlinecite{vurgaftman_band_2001}] or [\onlinecite{vurgaftman_band_2003}]).
For the II-VI bulk alloy Cd$_x$Zn$_{1-x}$Se for example,
a broad range of values between $b=0$ and $b=1.26$ \eV has been reported
throughout publications from the last two decades [\onlinecite{tit_absence_2009}] [\onlinecite{ammar_structural_2001}][\onlinecite{venugopal_photoluminescence_2006}]
[\onlinecite{gupta_optical_1995}]. The large disparity
on the experimental side can for example result from difficult growth conditions for the
mixed systems. On the theoretical side, the inadequate use of too simple approaches
like the VCA can lead to wrong results.

\begin{figure*}
    \centering
    \includegraphics[width=0.95\linewidth]{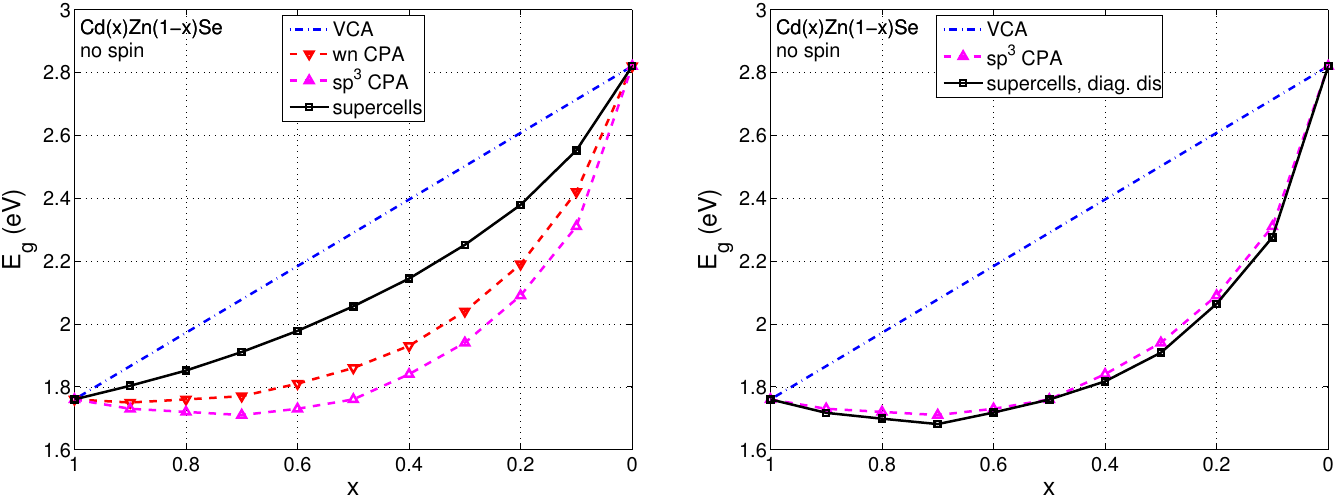}
    \caption{(Color online) Concentration-dependent band gap $E_{\text{g}}(x)$ of
             Cd$_x$Zn$_{1-x}$Se, calculated in the VCA (blue), $sp^3$ CPA (magenta),
             Wannier CPA (red, only in left figure)
             and supercell method (black). The spin-orbit coupling is 
             set to zero.
             In the left figure, the supercell results map the full disorder.
             In the right figure, the disorder in the 
             supercells  is artificially restricted to the site-diagonal elements
             in the $sp^3$ basis. See text for more details.}
    \label{fig:comparisonBowing}
\end{figure*}

In the CPA, the band gap can in principle be read off from the DOS. For numerical reasons,
the CPA DOS will not completely fall to zero in the bandgap, as a finite imaginary part 
$\delta$ of the energy is required. Nevertheless, it is possible to identify the band gap
with  desired accuracy by increasing the resolution. 
We used an imaginary part of $\delta = 10^{-4}$ \eV and $\#\f{k} \approx 10^7$
values in the irreducible BZ. The high BZ resolution has turned out
to be crucial to obtain convergence
for the results for the band gap. In the Wannier representation,
it is furthermore very important to carefully 
discretize the VCA DOS when using Eq.\,(\ref{eq:CPAGreenbanddiagonal})
because the DOS will contain kinks that stem from
Van Hove singularities [critical points
where $\operatorname{grad}_\f{k} E_n(\f{k})= 0$].
The usage of TB models that give a reliable band structure throughout the whole BZ
(in contrast to dispersions from  effective mass or $\f{k}\cdot\f{p}$ models)
will additionally lead to sharp peaks at some of these critical points, as the slope
of non-degenerate bands must also vanish at the BZ boundaries.

Figure \ref{fig:CPAbandedge}
shows the conduction band edge region of
Cd$_{0.5}$Zn$_{0.5}$Se, calculated in the Wannier CPA.
The peak of spectral function of the $\f{k}=\f{0}$
quasiparticle obviously coincides with the band edge within the chosen
energy resolution of $\Delta z = 5\times10^{-3}$ \eV. A further analysis
(not shown) reveals that the CPA band
edge states are mostly $\delta$-like; in case of broadened peaks, the corresponding 
linewidth is so small that the peak position still allows for a convenient
determination of the conduction band edge $E_{\text{c}}$ and the 
valence band edge $E_{\text{v}}$ with an accuracy of $0.01$ \eV. Hence, the
band gap $E_{\text{g}}(x) = E_{\text{c}}(x)- E_{\text{v}}(x)$ of the alloy can be
determined with an accuracy of 0.02 \eV.

The corresponding supercell band gap is given by
\begin{equation}
E_{\text{g}}(x) = \operatorname{min}\left\{ E_{\text{c}}^i(x) \right\} -
                  \operatorname{max}\left\{ E_{\text{v}}^i(x) \right\},
\end{equation} 
where $i$ again numbers the distinct configurations. Note that this definition of the 
band gap implies
\begin{equation}
E_{\text{g}}(x) \neq \frac{1}{N} \sum\limits_i
                      \left[E_\text{c}^i(x) - E_\text{v}^i(x)\right],
\end{equation}
i.\,e.\,the band gap of the disordered alloy is different from the configurational
average over the energy gaps for fixed realizations $i$. 

The resulting  curves for the Cd$_x$Zn$_{1-x}$Se
band gap are depicted in Fig.\,\ref{fig:comparisonBowing}.
In order to perform a detailed examination of the influence of the disorder and
the applicability of the CPA,
we performed two different supercell calculations for the Cd$_x$Zn$_{1-x}$Se alloy
system. In order to get rid of finite size effects, we used supercells with
$N_\f{R}=4000$ lattice sites and calculated the band edges for $N=50$ distinct
configurations per concentration. The concentration itself is varied in steps of 0.1.

The left subfigure
contains results from the supercell EBOM exactly as described in
Sec.\,\ref{subsec:supercellTBcalculation}, thus including the full
disorder in the site-diagonal and the hopping matrix elements on the microscopic
scale.  
The supercell results in the right subfigure have been obtained under the 
artificial restriction to site-diagonal disorder. This means
that only the site-diagonal matrix elements differ throughout
the cell and with each configuration, while the hopping matrix elements for
each concentration were substituted by their VCA values. Both the supercell 
approach and the CPA then use the same level of mean-field approximation for
the hopping matrix elements.
As each of supercell bowing curve requires the partial diagonalization
of $9 \times 50 = 450$ Hamiltonian matrices (the $x=0$ and $x=1$ values
are input material parameters), we additionally neglected the spin-orbit
coupling for this comparison. Besides an increase in computation time
by a factor $2^3=8$, the memory comsumption is lowered, as all matrix entries
are then real numbers.
 We also added the pure VCA results to both
figures---as the
tight-binding matrix is diagonal at $\f{k=0}$, we are left with a linear
$E_\text{g}(x)$ curve, which is at least able to indicate the deviation
of the other results from a linear interpolation. 

We will first turn to the \emph{full disorder} case on the left. Overall,
the bowing obtained in the CPA is clearly larger than the corresponding
supercell result. When the Wannier representation is used, the bowing curve is
slightly closer to the supercell case than in the $sp^3$ basis. This was to
be expected, as the overall shapes and bandwidths of the pure CdSe and ZnSe bands
are very much alike (see again Fig.\,\ref{fig:BS_DOS_CdSe_ZnSe}). Nevertheless,
the CPA (as well as the VCA) fails to satisfactorily reproduce the supercell
band gaps over the whole concentration range. We can even identify
a slight ``overbowing'', where the CPA band gap of the alloy dips beneath
the value for pure CdSe for the Cd-rich concentrations. This effect
is ultimately a consequence of the usage of the concentration-dependent VCA average
for the non-diagonal 
part of the Hamilton. Consequently, it does not occur in standard textbook examples,
where the same hopping values for the constituents are used and one is left
with a concentration-independent bandwidth.

A look at the right subfigure, where the supercell results with 
\emph{diagonal disorder} are given along with the $sp^3$ CPA curve, clearly reveals
the importance of the non-diagonal part (as the supercell potential
 lacks translational invariance, our supercell TB model
can only be used in the $sp^3$ basis, so that the Wannier CPA results
are also not given again here).
We easily notice that the CPA and supercell results now coincide very well.
Consequently, we can state that the CPA can in principle simulate the influence
of the disorder in the site-diagonal
elements on the band gap and the deviations stem from the mean-field
treatment for the hopping.
If we enforce the constraint 
$N_\text{AC}^i/ N_\f{R}^i=x$  for single concentration values
 and thus eliminate
the small influence of concentrational disorder (see Sec.
\ref{subsec:supercellTBcalculation}) on the supercell results, the discrepancy is even smaller
(not shown).

\begin{figure}
    \centering
    \includegraphics[width=0.95\linewidth]{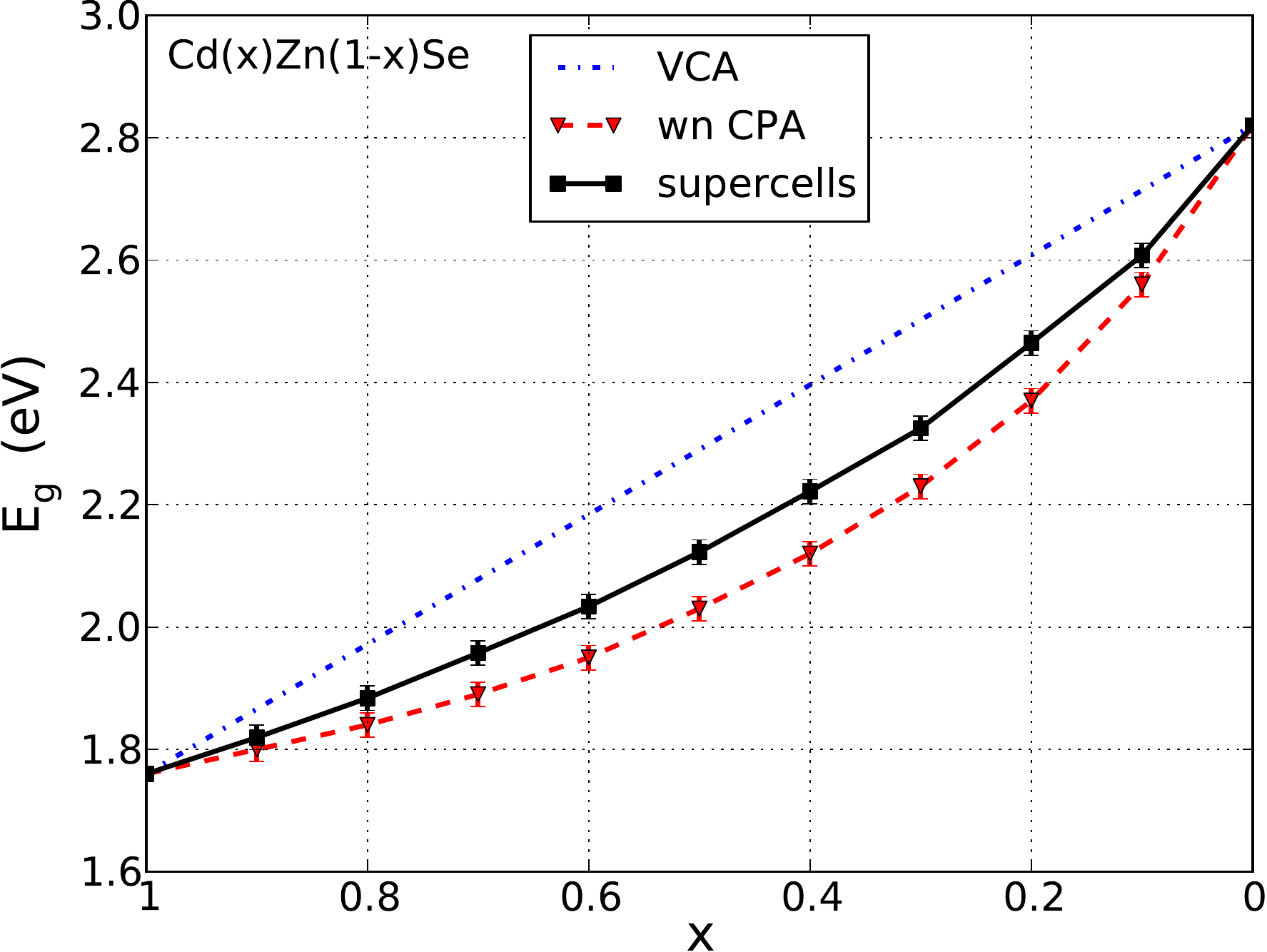}
    \caption{(Color online) Concentration-dependent band gap $E_{\text{g}}(x)$ of
             Cd$_x$Zn$_{1-x}$Se including spin-orbit coupling,
             calculated in the VCA (blue),
             Wannier CPA (red), and supercell method (black).}
    \label{fig:bowingCdZnSe}
\end{figure}

For the sake of completeness, we depict the Wannier CPA and supercell
results including  spin-orbit interaction in Fig.\,\ref{fig:bowingCdZnSe}
(the $sp^3$ CPA results will be omitted from now on).
We also added errorbars that account for the reading accuracy and finite size
effects. 
The difference in the first moments of the CdSe and ZnSe conduction bands turns out to be
slightly smaller when the spin is included. Obviously, the bowing is reduced and the 
deviation between the Wannier CPA and supercell results decreases further.
Still, only the $x \geq 0.9$ results overlap within the error range.
The best possible second-order fit to the $E_{\text{g}}(x)$ curve
yields a bowing parameter of $b=(0.71\pm0.09)$ \eV for the supercell results
and $b=(1.0\pm0.1)$ \eV for the curve as calculated with the Wannier CPA 
(note that the bowing values in [\onlinecite{mourad_band_2010}] were calculated for 
slightly differing  band gap values in the pure case).
It should be emphasized that the error range for the bowing values can only account
for the influence of the reading accuracy and finite size effects, as well
as for the deviation from a parabolic behaviour
(this is always to be expected for differing lattice
constants, see [\onlinecite{richardson_dielectric_1973}]).
The disparities that can arise from the uncertainty in the
input parameters for the pure materials (band gaps,
effective masses, valence band offsets) from different sources in the literature
cannot reliably be estimated at reasonable expense, so that 
the error range is certainly underestimated.

%==========================================================================
\subsection{Band gap bowing of In$_{x}$Ga$_{1-x}$As
\label{subsec:bowingInGaAs}}
%==========================================================================
%
\begin{figure*}
    \centering
    \includegraphics[width=0.95\linewidth]{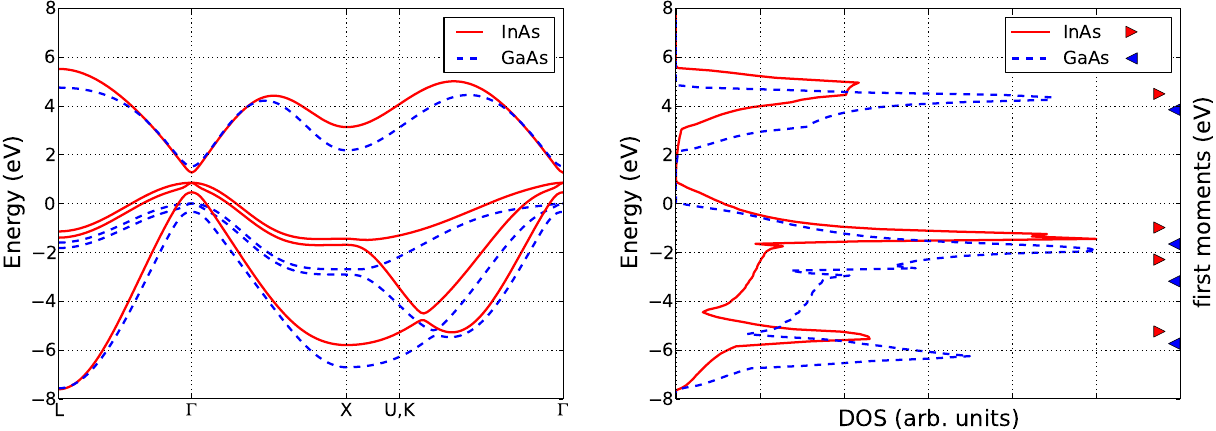}
    \caption{(Color online) Band structure (left) and DOS and first moments
             (right) of InAs and GaAs,
             calculated in the $sp^3$ EBOM.
             Again, the energies of InAs are shifted by a valence band offset
             $\Delta E_\text{vb} = 0.85$ \eV with respect to GaAs.
             See appendix for details.}
    \label{fig:BS_DOS_InAs_GaAs}
\end{figure*}

As a second example for the calculation of the band gap bowing, we apply the
methods to  
the zincblende bulk alloy In$_x$Ga$_{1-x}$As. The band structures, DOS and corresponding
first moments of InAs and GaAs are given in Fig.\,\ref{fig:BS_DOS_InAs_GaAs}. 
A valence band offset of $\Delta E_\text{vb} = 0.85$ \eV
[\onlinecite{pryor_eight-band_1998}] has already been incorporated.
This value has been obtained by the relative energetic position
to transition-metal impurities and is in agreement with experimental data
from measurements on Au Schottky barriers
[\onlinecite{tiwari_empirical_1992}].

\begin{figure}
    \centering
    \includegraphics[width=0.95\linewidth]{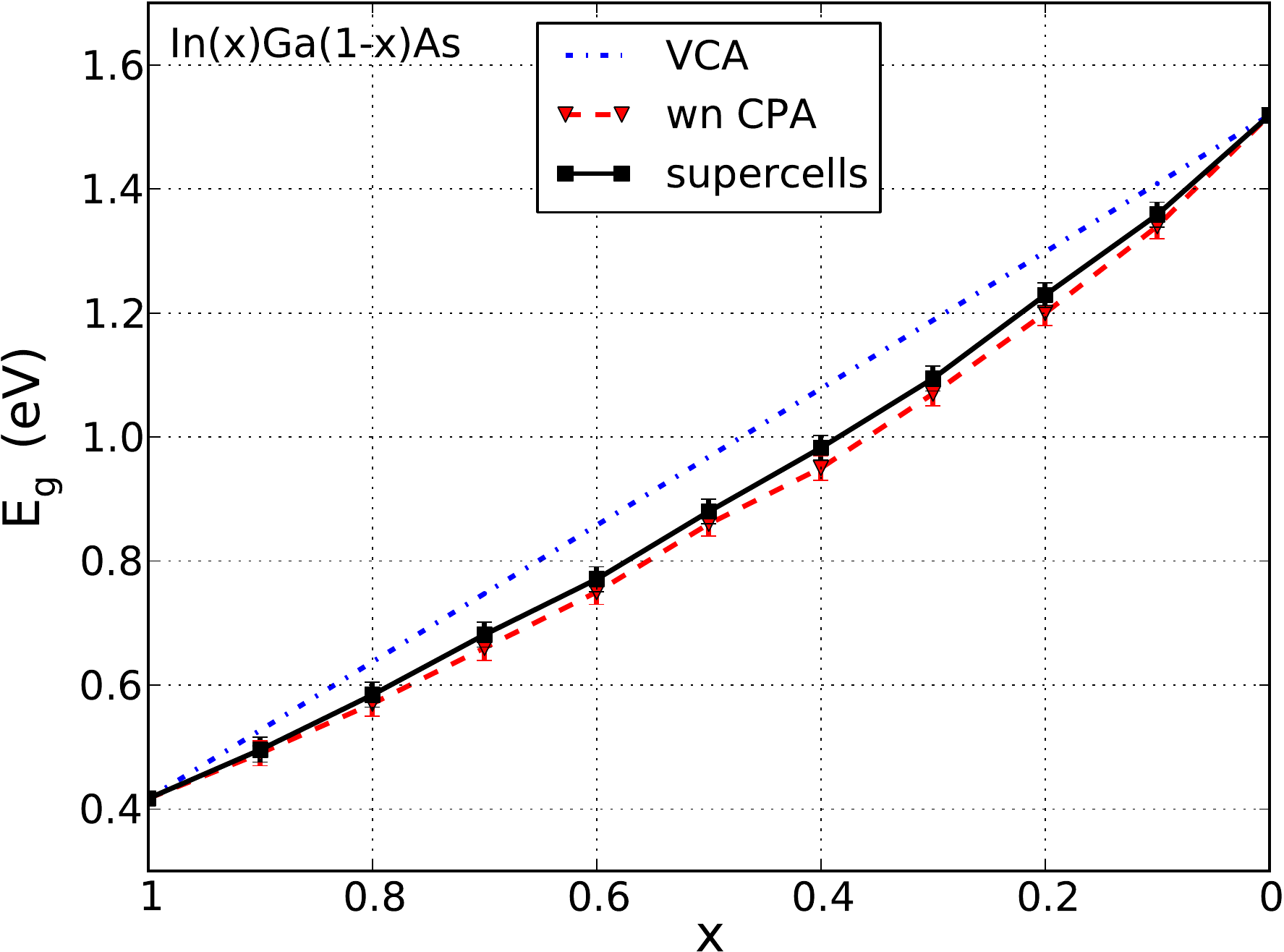}
    \caption{(Color online) Concentration-dependent band gap $E_{\text{g}}(x)$ of
             In$_x$Ga$_{1-x}$As including spin-orbit coupling,
             calculated in the VCA (blue), Wannier CPA (red)
             and supercell method (black).}
    \label{fig:bowingInGaAs}
\end{figure}

For this parameter set of InAs and GaAs the first moments of the 
conduction band are closer together than in the case of CdSe and
ZnSe, going along with a larger overlap of the bands.
Hence, we are closer to the weak scattering condition and therefore expect
better results for the concentration-dependent band gap from the (Wannier) CPA.
Again using $N=50$ configurations
and $N_\f{R}=4000$ lattice sites for the supercells and also the same
numerical parameters for the Wannier CPA, we obtain the bowing curve depicted
in Fig.\,\ref{fig:bowingInGaAs}.

Overall, the deviation from a linear behaviour (again indicated by the VCA results)
is smaller for In$_x$Ga$_{1-x}$As than for Cd$_x$Zn$_{1-x}$Se.
Furthermore, all results from the supercell calculations and the CPA 
calculations now overlap within the error range, although the 
CPA still slightly overestimates the bowing behaviour.
A second-order fit of the $E_\text{g}(x)$ values gives a bowing of
 $b=(0.39\pm0.04)$ \eV for the supercell results
and $b=(0.49\pm0.05)$ \eV for the curve as calculated with the Wannier CPA.
The fact that the resulting ranges touch
but do not overlap is mainly due to the deviation from a parabolic curve.
Nevertheless, including the error boundaries both values lie in the range
between $b=0.32$--$0.46$ \eV that is
recommended in the literature by Vurgaftman \etal in [\onlinecite{vurgaftman_band_2001}].
More recent \textit{ab initio} calculations within the DFT+LDA
(which are known to systematically underestimate the band gap) give a slightly
larger bowing in the range of $0.5$--$0.8$ \eV \cite{stroppa_composition_2005}.

%

%==========================================================================
\subsection{Band gap bowing of Ga$_{x}$Al$_{1-x}$N
\label{subsec:bowingGaAlN}}
%==========================================================================

\begin{figure*}
    \centering
    \includegraphics[width=0.95\linewidth]{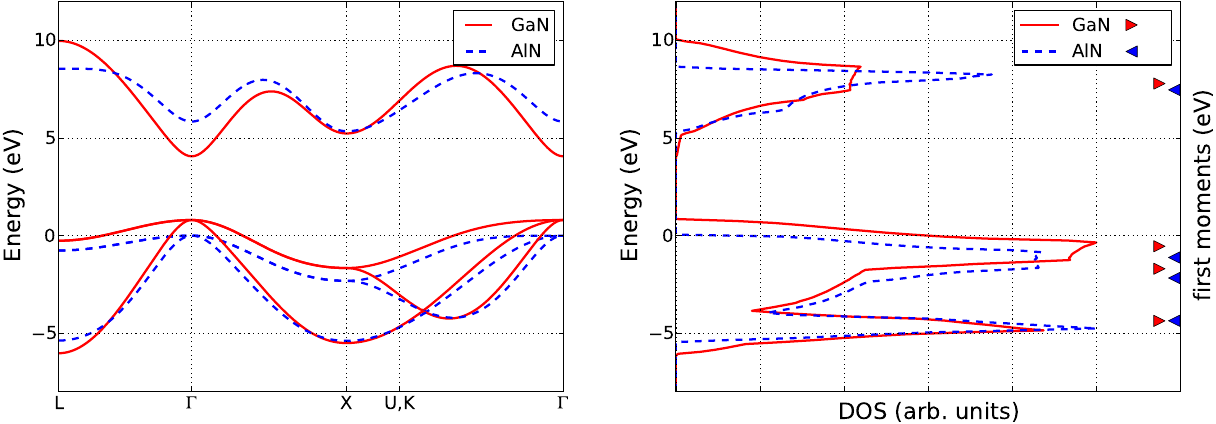}
    \caption{(Color online) Band structure (left) and DOS and first moments
             (right) of zincblende GaN and AlN,
             calculated in the $sp^3$ EBOM, assuming a valence band offset
             $\Delta E_\text{vb} = 0.8$ \eV and an indirect band gap for AlN.
             See appendix for details.}
    \label{fig:BS_DOS_GaN_AlN}
\end{figure*}

As a final example, we will calculate the concentration dependent band
gap of the zincblende phase of Ga$_{x}$Al$_{1-x}$N, which is frequently used
as barrier material in optoelectronics [\onlinecite{vurgaftman_band_2001}].
While the wurtzite modification of AlN is the only Al-containing
III-V semiconductor with a direct band gap, its zincblende modification is most
likely indirect with the conduction band minimum at the $X$-point and the
valence band maximum at the $\Gamma$-point [\onlinecite{fritsch_band-structure_2003}
(although a direct band gap is sometimes also assumed, see
e.\,g.\,[\onlinecite{fonoberov_excitonic_2003}).
As can be seen in Fig.\,\ref{fig:BS_DOS_GaN_AlN},
the $sp^3$ EBOM is able to reproduce the
indirect band gap of
AlN properly, as it is also fitted to the $X$-point energies. We use 
a valence band offset
 of $\Delta E_\text{vb} = 0.8$ \eV [\onlinecite{vurgaftman_band_2001}][\onlinecite{vurgaftman_band_2003}];
the resulting relative positions of the first moments and the large conduction band
overlap indicate again a good applicability of the CPA.
In contrast to the previous material systems,
the spin-orbit coupling is one order of magnitude smaller
in the nitride compounds and does not influence the results for the bowing.

\begin{figure}
    \centering
    \includegraphics[width=0.95\linewidth]{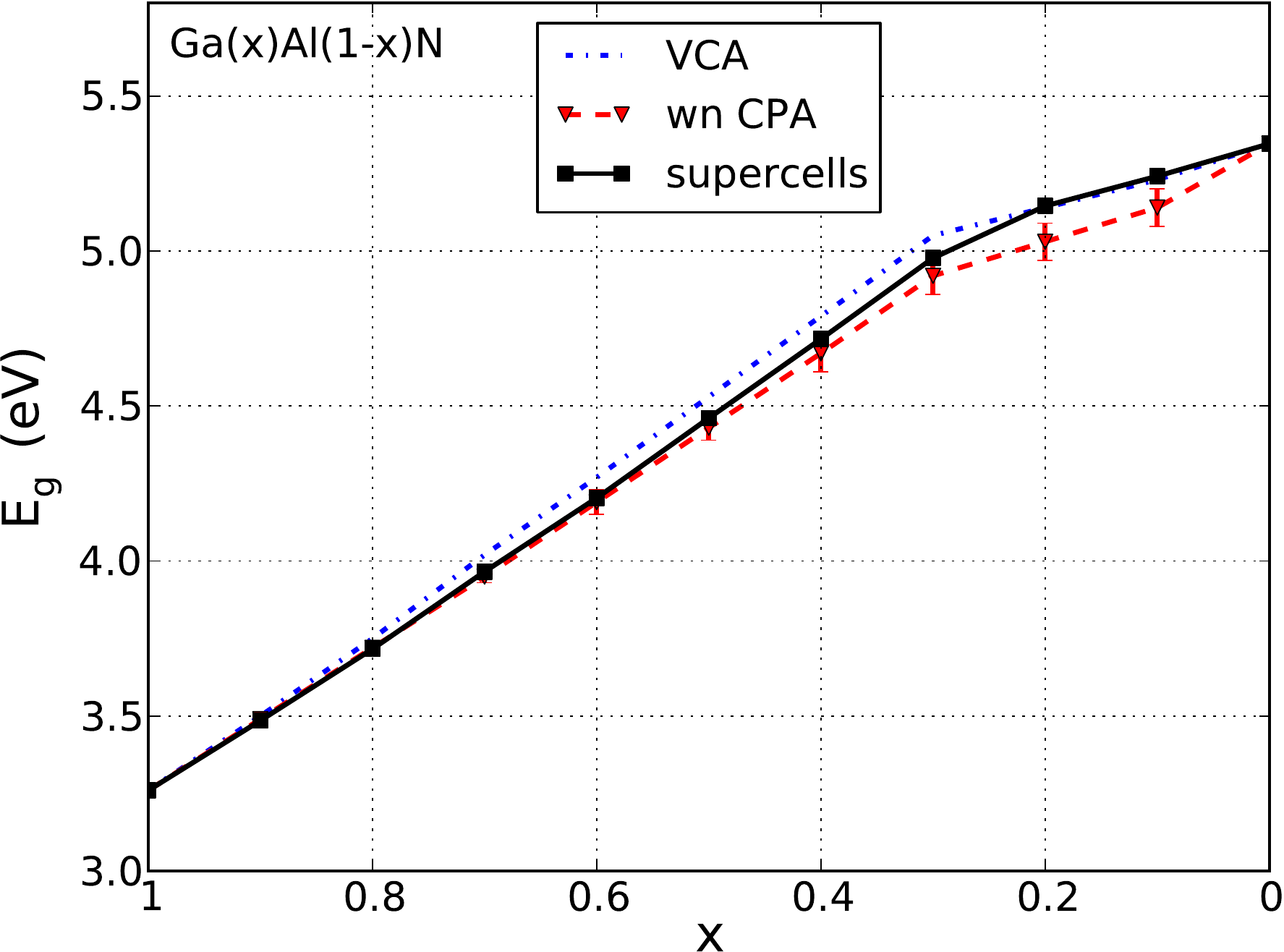}
    \caption{(Color online) Concentration-dependent band gap $E_{\text{g}}(x)$ of
             Ga$_x$Al$_{1-x}$N,
             calculated in the VCA (blue), Wannier CPA (red)
             and supercell method (black).}
    \label{fig:bowingGaAlN}
\end{figure}

The $E_{\text{g}}(x)$ results for Ga$_{x}$Al$_{1-x}$N 
are depicted in Fig.\,\ref{fig:bowingGaAlN}.
The numerical parameters for the CPA and the supercell calculations were chosen identical
to those of the previous sections.
In all three models, we can clearly identify 
a crossover in the bowing between $x=0.3$ and $x=0.2$. In this region,
the character of the band gap changes from a behaviour which is strongly influenced
by the indirect AlN material to a direct band gap behaviour dominated by
GaN.

While the CPA results coincide  very well with the supercell results
for the GaN-dominated side at large $x$, they overestimate the bowing
on the Al-rich side. In contrast, the VCA, which  shows a piecewise linear
behaviour with two different slopes, can reproduce the supercell
results for high Al contents quite well, but deviates for $0.7 \geq x \geq 0.3$.
It additionally should be noted that the CPA results in the Al-rich part
have larger error bars, as the determination of the band gap
is afflicted with a larger uncertainty in this range due to a larger 
broadening of the corresponding spectral functions (not shown).

In [\onlinecite{vurgaftman_band_2003}], Vurgaftman \etal report
bowing parameters of $b_\Gamma= 0.05$--$0.53$ \eV for the $\Gamma$-valley
of cubic Ga$_{x}$Al$_{1-x}$N from theory. By additionally taking 
several experimental
results into account (which obviously render larger bowing parameters),
they recommend an approximate value 
of $b_\Gamma\approx0.7$ \eV. By only fitting the $\Gamma$-valley bowing,
i.\,e.\, only taking 
the values for $x\geq0.3$ into account, augmented by the AlN energy difference
at $\Gamma$ as boundary value, we obtain bowing parameters of 
$b_\Gamma = (0.37\pm0.02)$ \eV from the supercell calculations and   
$b_\Gamma = (0.4\pm0.1)$ \eV from the CPA. Consequently, the CPA
and supercell values agree within the error range and our results are in 
reasonably good agreement with the literature values. More recent
results from 
DFT+LDA calculations [\onlinecite{kanoun_ab_2005}] yield a value
of $b_\Gamma \approx 0.5$ \eV and predict the crossover at about $x=0.4$.
However, it should be noted that these results suffer from the usual
underestimation of band gaps in DFT+LDA. The pure GaN and AlN band gaps in
their calculations are obtained as 1.93 \eV and 3.23 \eV respectively.
Consequently, they
strongly deviate from our input values of 3.26 \eV for GaN and 5.346 \eV
for AlN (see appendix), as our CPA/supercell + EBOM model allows for
the usage of arbitrarily exact boundary values at $x=0$ and $x=1$.

For the sake of comparison,
we summed up the results for the bowing parameters  in Tab.\,\ref{tab:table_bowing}.
As already stated, the given error ranges only account for the reading accuracy,
finite size effects and non-parabolicity of $E_\text{g}(x)$ but cannot reflect 
the reliability of the input band structure parameters for the pure materials.  

\begin{table}
\centering
\caption{Summary of the results for the bowing parameter $b$ in \eV.
         In case of Ga$_x$Al$_{1-x}$N, the value solely refers to the bowing
         of the $\Gamma$-valley.}
\label{tab:table_bowing}
\begin{tabularx}{0.95\linewidth}{c|ccc}
 Material &  Supercells & Wannier CPA & Literature \\
\hline
\hline
 Cd$_x$Zn$_{1-x}$Se  & $0.71\pm0.09$ & $1.0\pm0.1$ & $0$--$1.26$ \\
 In$_x$Ga$_{1-x}$As  & $0.39\pm0.04$ & $0.49\pm0.05$ & $0.32$--$0.46$\\
 Ga$_x$Al$_{1-x}$N   & $0.37\pm0.02$ & $0.4\pm0.1$ & $0.05$--$0.7$
\end{tabularx}
\end{table}

%==========================================================================
%                                 Conclusion
%==========================================================================
\section{Conclusion and outlook \label{sec:conclusion}}

In this paper, we showed that the electronic properties of substitutional
semiconductor alloys with an underlying
zincblende structure of the type A$_x$B$_{1-x}$C can well be described with 
empirical TB models.
We presented a combined theoretical approach, starting from a multiband TB
model with a realistic dispersion and bandwidth throughout the whole Brillouin zone.
The density of states and the band gap
of the disordered system can then either be determined
by exact diagonalization of a large supercell with
a large number $N$ of microscopically
distinct configurations, or by combination
of the TB model with the  coherent potential approximation (CPA) and/or the 
virtual crystal approximation (VCA).

Using the supercell results for
Cd$_x$Zn$_{1-x}$Se as reference,
we gave a  careful quantitative and qualitative analysis of the scope
of validity of the CPA and VCA, especially with regards to the calculation
of the concentration dependent band gap $E_\text{g}(x)$ of the alloy.
While the VCA failed over the whole concentration range,
the CPA also turned out to be not accurate enough for the Cd$_x$Zn$_{1-x}$Se system under
consideration, although the proper choice of the basis set could significantly 
reduce the discrepancy.

We then applied our TB model to two further different alloy systems,
namely the III-V alloy In$_x$Ga$_{1-x}$As and the III-nitride system Ga$_x$Al$_{1-x}$N.
For both systems, the CPA gave good results.
In case of In$_x$Ga$_{1-x}$As, the CPA and the supercell calculations
yielded bowing parameters
in  good agreement with literature values from experiments.
 For  Ga$_x$Al$_{1-x}$N the band gap bowing showed
a crossover behaviour between $x=0.3$ and $x=0.2$,
due to the fact that cubic GaN has a direct energy gap at the BZ center,
while the conduction band minimum of cubic AlN is located at $X$. 
This crossover was reproduced in all three models. The CPA and the supercell 
calculations were able to reproduce the $\Gamma$-valley bowing in satisfactory
agreement with the literature. On the Al-rich side ($x\leq 0.3$), the CPA understimated
the band gap when compared to the supercell approach, while the VCA was in
surprisingly good agreement.

For the sake of completeness,
it should be emphasized that the computational costs of the CPA are
far smaller than in the supercell case. If the number of bands has to be
augmented or properties
far from the band edges become relevant, the supercell approach can quickly
become infeasible, as the calculation time scales with the cube of the dimension
of the Hamiltonian matrix.

% The accuracy of the CPA can in principle be augmented by the inclusion
% of off-diagonal disorder into the CPA scheme [\onlinecite{matsubara_structure_1982}, 
% at the cost of losing the simplicity of the CPA iterative scheme.

As the supercell calculations and, under certain conditions outlined in this paper,
also the CPA can give good results for the concentration-dependent band gap when
combined with the ETBM and especially the EBOM,
the application to further material systems (with disorder either in
the cations or in the anions)
will be an interesting task for the future,
ideally alongside actual experimental data.
Furthermore, the same calculation scheme can be transferred to alloy systems
with  underlying wurtzite structure by using a suitable TB Hamiltonian
[\onlinecite{mourad_multiband_2010}] for direct as well as indirect band gap materials. 

Our supercell approach is also applicable to disordered low-dimensional structures,
as shown in [\onlinecite{mourad_band_2010}],[\onlinecite{mourad_multiband_2010-1}].
Therefore, in principle also disordered nanowires or superlattices
can be investigated. As the CPA is exact only in the limit of infinite 
dimensions, it may turn out that the CPA + EBOM is less reliable when applied
to low-dimensional systems, so that suitable extensions of the CPA must be used.

%==========================================================================
%                                 Appendix
%==========================================================================
\newpage 

\section*{Appendix \label{sec:appendix}}

\begin{appendix}

The following tables give the material parameters used throughout this
paper, including the  sources from the literature.

The band structures in the $sp^3$ EBOM can be fitted to the conventional
 lattice constant $a$,
the spin-orbit splitting
$\Delta_\text{so}$, the effective conduction band mass $m_\text{c}$, the Luttinger 
parameters $\gamma_1, \gamma_2, \gamma_3$, 
and to a set of energies at the $\Gamma$-point and 
the $X$-point (denoted by the usual single group notation),
where the energy gap $E_g$ is given as $\Gamma_1^\text{c}-\Gamma_{15}^\text{v}$
in case of a direct band gap.
Additionally, a valence band offset $\Delta E_\text{vb}$ is incorporated in 
order to account for the relative energetic position of the two constituents.

These parameters are uniquely connected to the non-vanishing TB matrix elements from
Eq.\,(\ref{eq:EBOMme}) of the present paper
by the equations $(7)$--$(17)$, $(33)$ and $(45)$--$(53)$
of Ref.\,\onlinecite{loehr_improved_1994}.
% \begin{acknowledgement}
%
% \end{acknowledgement}

\section{Material parameters  CdSe and ZnSe}

%
% \begin{table}
% \caption{Material parameters  for CdSe and ZnSe including sources.}
% \label{tab:MaterialparameterCdSeZnSe}
\begin{center}
\begin{tabularx}{0.9\linewidth}{lr|lr|lr}
Parameter & {} & CdSe & & ZnSe & \\

\hline
\hline
$a$ & (\AA) &  \phantom{+}6.078  & [\onlinecite{kim_optical_1994}]&  \phantom{+}5.668 &[\onlinecite{kim_optical_1994}]\\
$\Delta_\text{so}$ & (\eV) &  \phantom{+}0.41 &[\onlinecite{kim_optical_1994}] &  \phantom{+}0.43 &[\onlinecite{kim_optical_1994}]\\
$m_\text{c}$ & ($m_0$) &  \phantom{+}0.12  &[\onlinecite{kim_optical_1994}] &  \phantom{+}0.147 &[\onlinecite{hoelscher_investigation_1985}]\\
$\Gamma_1^\text{c}-\Gamma_{15}^\text{v}$  & (\eV) &  \phantom{+}1.76  &[\onlinecite{adachi_cubic_2004}]&  \phantom{+}2.82 &[\onlinecite{kim_optical_1994}]\\
$X_1^\text{c}$ & (\eV) &  \phantom{+}2.94 & [\onlinecite{blachnik_numerical_1999}]&  \phantom{+}4.41 &[\onlinecite{blachnik_numerical_1999}]\\
$X_5^\text{v}$ & (\eV) & $-$1.98 & [\onlinecite{blachnik_numerical_1999}]&$-$2.08 &[\onlinecite{blachnik_numerical_1999}]\\
$X_3^\text{v}$ & (\eV) & $-$4.28 & [\onlinecite{blachnik_numerical_1999}]&$-$5.03 &[\onlinecite{blachnik_numerical_1999}] \\
$\gamma_1$ &{}&  \phantom{+}3.33  &[\onlinecite{kim_optical_1994}] & \phantom{+}2.45 &[\onlinecite{hoelscher_investigation_1985}]\\
$\gamma_2$ &{}& \phantom{+}1.11  & [\onlinecite{kim_optical_1994}]&  \phantom{+}0.61 & [\onlinecite{hoelscher_investigation_1985}]\\
$\gamma_3$ &{}&  \phantom{+}1.45  & [\onlinecite{kim_optical_1994}]&  \phantom{+}1.11 &[\onlinecite{hoelscher_investigation_1985}]\\
$\Delta E_\text{vb}$ & (\eV) & \phantom{+}0.22  &[\onlinecite{kim_optical_1994}] &  \phantom{+}0 & [\onlinecite{kim_optical_1994}]
\end{tabularx}
\end{center}
% \end{table}

%
\section{Material parameters  InAs and GaAs}

% \begin{table}
% \caption{Material parameters  for InAs and GaAs including sources.}
% \label{tab:MaterialparameterInAsGaAs}
% \centering
\begin{center}
\begin{tabularx}{0.9\linewidth}{lr|lr|lr}
Parameter & {} & InAs & & GaAs & \\
\hline
\hline
$a$ & (\AA) &  \phantom{+}6.058   &[\onlinecite{loehr_inasn/inxga1xsbm_1995}]& \phantom{+}5.653 &[\onlinecite{vurgaftman_band_2001}]\\
$\Delta_\text{so}$ & (\eV) &  \phantom{+}0.39 &[\onlinecite{vurgaftman_band_2001}] &  \phantom{+}0.34 &[\onlinecite{vurgaftman_band_2001}]\\
$m_\text{c}$ & ($m_0$) & \phantom{+}0.022  &[\onlinecite{loehr_inasn/inxga1xsbm_1995}] & \phantom{+}0.067 &[\onlinecite{vurgaftman_band_2001}]\\
$\Gamma_1^\text{c}-\Gamma_{15}^\text{v}$  & (\eV) & \phantom{+}0.417  &[\onlinecite{vurgaftman_band_2001}]& \phantom{+}1.519 &[\onlinecite{vurgaftman_band_2001}]\\
$X_1^\text{c}$ & (\eV) & \phantom{+}2.28  &[\onlinecite{loehr_inasn/inxga1xsbm_1995}]& \phantom{+}2.18 &[\onlinecite{adachi_numerical_2002}]\\
$X_5^\text{v}$ & (\eV) & $-$2.42  &[\onlinecite{loehr_inasn/inxga1xsbm_1995}]&$-$2.8 &[\onlinecite{adachi_numerical_2002}]\\
$X_3^\text{v}$ & (\eV) & $-$6.64  &[\onlinecite{loehr_inasn/inxga1xsbm_1995}]&$-$6.7 &[\onlinecite{adachi_numerical_2002}] \\
$\gamma_1$ &{}& \phantom{+}20.0  &[\onlinecite{vurgaftman_band_2001}] &\phantom{+}6.98 &[\onlinecite{vurgaftman_band_2001}]\\
$\gamma_2$ &{}& \phantom{+}8.5   &[\onlinecite{vurgaftman_band_2001}]& \phantom{+}2.06 &[\onlinecite{vurgaftman_band_2001}]\\
$\gamma_3$ &{}& \phantom{+}9.2   &[\onlinecite{vurgaftman_band_2001}]& \phantom{+}2.93 &[\onlinecite{vurgaftman_band_2001}]\\
$\Delta E_\text{vb}$ & (\eV) & \phantom{+}0.85  &[\onlinecite{pryor_eight-band_1998}] & \phantom{+}0 &[\onlinecite{pryor_eight-band_1998}]
\end{tabularx}
\end{center}
% \end{table}

%
% \begin{table}
% \caption{Material parameters  for GaN and alN including sources.}
% \label{tab:MaterialparameterCdSeZnSe}
% \centering

\section{Material parameters  GaN and AlN}

\begin{center}
\begin{tabularx}{0.9\linewidth}{lr|lr|lr}
Parameter & {} & GaN & & AlN & \\
\hline
\hline
$a$ & (\AA) &  \phantom{+}4.50   &[\onlinecite{fonoberov_excitonic_2003}]&  \phantom{+}4.38 &[\onlinecite{fonoberov_excitonic_2003}]\\
$\Delta_\text{so}$ & (\eV) &  \phantom{+}0.017 &[\onlinecite{fonoberov_excitonic_2003}] &  \phantom{+}0.019 &[\onlinecite{fonoberov_excitonic_2003}]\\
$m_\text{c}$ & ($m_0$) &  \phantom{+}0.15  &[\onlinecite{fonoberov_excitonic_2003}] &  \phantom{+}0.25 &[\onlinecite{fonoberov_excitonic_2003}]\\
$\Gamma_1^\text{c}-\Gamma_{15}^\text{v}$  & (\eV) &  \phantom{+}3.26  &[\onlinecite{fonoberov_excitonic_2003}]& \phantom{+} 5.84 &[\onlinecite{fritsch_band-structure_2003}]\\
$X_1^\text{c}$ & (\eV) &  \phantom{+}4.43  &[\onlinecite{fritsch_band-structure_2003}]&  \phantom{+}5.346 &[\onlinecite{fritsch_band-structure_2003}]\\
$X_5^\text{v}$ & (\eV) & $-$2.46  &[\onlinecite{fritsch_band-structure_2003}]& $-$2.315 &[\onlinecite{fritsch_band-structure_2003}]\\
$X_3^\text{v}$ & (\eV) & $-$6.30  &[\onlinecite{fritsch_band-structure_2003}]&$-$5.388 &[\onlinecite{fritsch_band-structure_2003}] \\
$\gamma_1$ &{}&  \phantom{+}2.67  &[\onlinecite{fonoberov_excitonic_2003}] & \phantom{+}1.92 &[\onlinecite{fonoberov_excitonic_2003}]\\
$\gamma_2$ &{}&  \phantom{+}0.75   &[\onlinecite{fonoberov_excitonic_2003}]&  \phantom{+}0.47 &[\onlinecite{fonoberov_excitonic_2003}]\\
$\gamma_3$ &{}&  \phantom{+}1.10   &[\onlinecite{fonoberov_excitonic_2003}]&  \phantom{+}0.8$^*$\\
$\Delta E_\text{vb}$ & (\eV) &  \phantom{+}0.8  &[\onlinecite{vurgaftman_band_2003}] &  \phantom{+}0 &[\onlinecite{fonoberov_excitonic_2003}]
\end{tabularx}
\end{center}
$^*$This parameter has been adjusted by hand, as the original value
of $\gamma_3$=0.85 used in [\onlinecite{fonoberov_excitonic_2003}] leads to an
erroneous curvature in the $\Gamma$--$K$ direction.
% \end{table}

\end{appendix}

%==========================================================================
%                                bibliography
%========================================================================== 

% 
% \bibliography{./MeineBibliothek}
% \bibliographystyle{epj}
%\bibliographystyle{apsrev}
% \begin{thebibliography}{88}
% 

\end{document}